\documentclass[sigconf]{acmart}
\usepackage{xspace}
\usepackage{multirow}
\usepackage{booktabs}       
\usepackage{array}          
\usepackage{colortbl}
\usepackage[normalem]{ulem} 
\usepackage{makecell}
\usepackage{listings}

\definecolor{mp}{HTML}{8AB174} 
\definecolor{sp}{HTML}{B16F6C} 
\definecolor{np}{HTML}{A1A1A1} 

\newcommand{\hlcell}[3][100]{%
  \begingroup
  \setlength{\fboxsep}{0pt}
  \vphantom{\rule{0pt}{3ex}}
  \colorbox{#2!#1}{\makebox[3em][c]{\strut #3}}
  \endgroup
}

\definecolor{color0}{rgb}      {0.0,0.0,0.0}
\definecolor{color1}{rgb}      {0.9,0.5,0.0}
\definecolor{color2}{rgb}      {0.6,0.0,0.8}
\definecolor{color3}{rgb}      {0.0,0.4,0.5}
\definecolor{color4}{rgb}      {0.8,0.3,0.5}
\definecolor{color5}{rgb}      {0.9,0.0,0.0}
\definecolor{color6}{rgb}      {0.0,0.6,0.8}
\definecolor{color7}{rgb}      {0.5,0.4,0.0}
\definecolor{color14}{rgb}      {1.0,0.0,0.0}

\definecolor{color8}{HTML}     {417DD7}
\definecolor{color9}{HTML}     {E85C6C}
\definecolor{color10}{HTML}    {EEBA33}
\definecolor{color11}{HTML}    {65C079}
\definecolor{color12}{HTML}    {B172CF}
\definecolor{color13}{HTML}    {8AB174}

\definecolor{darkgrey-01}{HTML}    {4c4c4f}
\definecolor{darkgrey-02}{HTML}    {2a2a2e}

\newcommand{\reviseY}[1]{\textcolor{color0}{#1}}

\newcommand{\revise}[1]{\textcolor{color0}{#1}}

\newcommand{\Condition}[1]{\textsc{#1 condition\xspace}}

\newcommand{\conditionM}{\Condition{multi}\xspace}
\newcommand{\conditionS}{\Condition{single}\xspace}
\newcommand{\conditionN}{\Condition{none}\xspace}
\newcommand{\condM}{\textsc{multi}\xspace}
\newcommand{\condS}{\textsc{single}\xspace}
\newcommand{\condN}{\textsc{none}\xspace}

\usepackage{amsmath}
\usepackage{graphicx}
\usepackage{here}
\PassOptionsToPackage{hyphens}{url}
\usepackage{hyperref} 
\usepackage{url} 
\usepackage{textcomp} 
\usepackage{multicol}
\usepackage{subcaption}
\usepackage{comment}
\usepackage{booktabs,dcolumn}
\usepackage{multirow}
\usepackage{fancyhdr}			

\renewenvironment{quote}[1][0.04\linewidth]
  {\list{}{\leftmargin=#1\rightmargin=#1}\item\relax}{\endlist}

\usepackage{color}
\definecolor{orange}{RGB}{255,127,0}
\definecolor{limegreen}{RGB}{50, 205, 50}
\definecolor{violet}{RGB}{148,0,211}

\newcolumntype{L}[1]{>{\raggedright\let\newline\\\arraybackslash\hspace{0pt}}m{#1}}
\newcolumntype{C}[1]{>{\centering\let\newline\\\arraybackslash\hspace{0pt}}m{#1}}
\newcolumntype{R}[1]{>{\raggedleft\let\newline\\\arraybackslash\hspace{0pt}}m{#1}}

\newif\ifCOMMENTS
\COMMENTStrue

\ifCOMMENTS

\else

\fi

\ifCOMMENTS

\else

\fi
\newcommand{\RQtwo}{How do thought exchanges with AI agents, compared with reading solely, influence readers’ critical-thinking skills?}
\newcommand{\RQthree}{How do single-agent and multi-agent thought exchanges shape readers’ critical reading practices and interaction strategies? }

\makeatletter
\def\Hline{
  \noalign{\ifnum0=`}\fi\hrule \@height 4.\arrayrulewidth \futurelet
   \reserved@a\@xhline}
\makeatother
\AtBeginDocument{%
  }

\setcopyright{acmlicensed}

\copyrightyear{2026}
\acmYear{2026}
\setcopyright{cc}
\setcctype{by}
\acmConference[IUI '26]{31st International Conference on Intelligent User Interfaces}{March 23--26, 2026}{Paphos, Cyprus}
\acmBooktitle{31st International Conference on Intelligent User Interfaces (IUI '26), March 23--26, 2026, Paphos, Cyprus}
\acmPrice{}
\acmDOI{10.1145/3742413.3789069}
\acmISBN{979-8-4007-1984-4/2026/03}

\begin{document}

\title{LLM-based In-situ Thought Exchanges for Critical Paper Reading}

\author{Xinrui Fang}
\email{xinrui.fang@iis-lab.org}
\orcid{0000-0001-5444-8722}
\affiliation{%
  \institution{IIS Lab, The University of Tokyo}
  \city{Tokyo}
  \country{Japan}
}

\author{Anran Xu}
\email{anran.xu@riken.jp}
\orcid{0000-0003-1790-7859}
\affiliation{%
  \institution{RIKEN Center for Advanced Intelligence Project, RIKEN}
  \city{Tokyo}
  \country{Japan}
}
\authornote{Also with IIS Lab, The University of Tokyo.}

\author{Chi-Lan Yang}
\email{chilan.yang@iii.u-tokyo.ac.jp}
\orcid{0000-0003-0603-2807}
\affiliation{%
  \institution{IIS Lab, The University of Tokyo}
  \city{Tokyo}
  \country{Japan}
}

\author{Ya-Fang Lin}
\email{yafanglin@iis-lab.org}
\orcid{0000-0003-3689-5910}
\affiliation{%
  \institution{IIS Lab, The University of Tokyo}
  \city{Tokyo}
  \country{Japan}
}

\author{Sylvain Malacria}
\email{sylvain.malacria@inria.fr}
\orcid{0000-0002-5201-5875}
\affiliation{%
  \institution{Univ. Lille, Inria, CNRS, Centrale Lille, UMR 9189 CRIStAL}
  \city{Lille}
  \country{France}
}
\authornotemark[1]

\author{Koji Yatani}
\email{koji@iis-lab.org}
\orcid{0000-0003-4192-0420}
\affiliation{%
  \institution{IIS Lab, The University of Tokyo}
  \city{Tokyo}
  \country{Japan}
}

\renewcommand{\shortauthors}{Fang et al.}




\begin{abstract}
Critical reading is a primary way through which researchers develop their critical thinking skills.
While exchanging thoughts and opinions with peers can strengthen critical reading, junior researchers often lack access to peers who can offer diverse perspectives.
To address this gap, we designed an in-situ thought exchange interface informed by peer feedback from a formative study (N=8) to support junior researchers’ critical paper reading.
We evaluated the effects of thought exchanges under three conditions (no-agent, single-agent, and multi-agent) with 46 junior researchers over two weeks.
Our results showed that incorporating agent-mediated thought exchanges during paper reading significantly improved participants’ critical thinking scores compared to the no-agent condition.
In the single-agent condition, participants more frequently made reflective annotations on the paper content.
In the multi-agent condition, participants engaged more actively with agents’ responses.
Our qualitative analysis further revealed that participants compared and analyzed multiple perspectives in the multi-agent condition.
This work contributes to understanding in-situ AI-based support for critical paper reading through thought exchanges and offers design implications for future research.

\end{abstract}  
\maketitle 


\section{Introduction}

Reading academic papers requires critical thinking and is more than a solitary activity.
Peer discussion helps readers develop a more comprehensive understanding of a paper’s content and a balanced view of its contributions, limitations, and implications~\cite{keller1993comprehensive}. 
Such thought exchanges, which involve articulating, questioning, and refining ideas through dialogue with others, help readers clarify difficult concepts, confront different arguments, and think of alternative interpretations that may otherwise be ignored.
However, it is often difficult to find peers reading the same paper or working on similar topics because researchers’ schedules and interests do not always align.
As a result, opportunities for meaningful peer discussion are often limited, leaving individuals to interpret academic papers in isolation.

A potential solution involves using conversational agents enabled by large language models (LLMs). Rather than relying solely on human peers, readers may seek out these agents as discussion partners.
Previous studies have shown that interacting with LLM-enabled agents can enhance comprehension~\cite{Schmuker2024Ruffle}, stimulate critical thinking~\cite{Tanprasert2024Debate}, and offer multiple perspectives~\cite{SeeWidely2024}.
Recent studies further suggest that the number of agents involved differently influences people’s cognitive processing.
For instance, single-agent settings are often perceived as more usable~\cite{clarke2024one}, but they may keep users stay engaged with a single line of thought.
Alternatively, multi-agent settings more accurately simulate the dynamics of group discussion by presenting competing viewpoints~\cite{park2023choicemates, SeeWidely2024}, raising counterarguments~\cite{liang2023encouraging}, and providing diverse lines of reasoning~\cite{liu2025breaking}.
This suggests that both single- and multi-agent approaches could extend the intellectual benefits of peer discussion and may usefully support critical reading of academic papers.

Although simulating agents as discussion peers has the potential to enhance critical paper reading, there is a lack of understanding about how different formats of agent-enabled discussions influence the reading experience.
In particular, it is unclear whether readers benefit more from one-on-one conversations with a single agent or from simulated group thought exchange where multiple agents contribute diverse viewpoints.
To clarify this, we investigate the following research questions:

\textbf{RQ1:} \RQtwo

\textbf{RQ2:} \RQthree

To answer these questions, we first conducted a formative study with eight participants, whose results helped us identify design goals for facilitating critical paper reading.
These goals informed the design of an LLM-based in-situ thought exchange interface to support critical reading. 
We then conducted a two-week user study with 46 junior researchers to examine how no-agent, single-agent, and multi-agent formats influenced critical reading and thinking. 

The results showed that both single- and multi-agent thought exchanges significantly improved participants’ critical-thinking scores compared with the no-agent condition.
Although the single- and multi-agent formats did not differ significantly in critical-reading scores, they shaped readers’ practices differently. 
The single-agent interface encouraged reflective annotation upon receiving responses, whereas the multi-agent interface prompted readers to devote more time to analyzing perspectives across agents.
Furthermore, open-ended questions analysis suggested that interacting with multiple agents may enhance participants’ analytical skills compared to single-agent formats, but it can also lead to feelings of being overwhelmed by too much external information.


\section{Related Work}

\subsection{Multi-perspective and interdisciplinary approaches in critical thinking}
\label{sec: related work 1}
Critical thinking has long been recognized as a fundamental skill in education~\cite{abrami2008instructional, bailin2003critical, dekker2020teaching} and academic paper reading~\cite{abroart,coiro2021toward}, encompassing higher-order abilities such as analysis, synthesis, and evaluation in Bloom’s Taxonomy~\cite{bloom1956taxonomy, Soozandehfar2016ACA, o2021critical, lee2025impact}.

Existing research has examined how incorporating multiple perspectives into learning and reading activities can help learners develop critical-thinking abilities~\cite{abrami2008instructional, dekker2020teaching, shihab2011reading, French2014DetectionOD}.
Dekker et al.~\cite{dekker2020teaching} argued that critical thinking entails questioning absolute knowledge claims and recognizing that problems can be addressed from multiple scientific and social perspectives.
Similarly, Shihab~\cite{shihab2011reading} showed that readers from different backgrounds interpret the same text differently, underscoring the value of exposure to multiple perspectives.
One widely used way to foster critical thinking through multiple viewpoints is group discussion among participants with diverse perspectives~\cite{kamin2002does,perry1999forms}.
In particular, Tsui~\cite{tsui2002fostering} found that group discussions can improve students’ self-reported critical thinking abilities.
Terenzini et al.~\cite{terenzini1995influences} also showed that collaborative learning—exchanging ideas with peers, mentors, or instructors—plays a central role in developing students’ critical-thinking skills.
However, Yuan et al.~\cite{Yuan2023CriTrainer} noted that coordinating group discussions demands substantial effort, limiting opportunities for early-career researchers to practice critical paper reading.

In research contexts, integrating multiple perspectives can be achieved by drawing on knowledge from different domains. 
Human knowledge is traditionally classified into distinct disciplines~\cite{abbott2010chaos}, but there is a growing trend toward interdisciplinary study approaches~\cite{van2015interdisciplinary}. 
In response, new tools have emerged to support interdisciplinary research and help researchers generate perspectives that integrate knowledge across domains~\cite{Zheng2024DiscipLink, Liu2024Selenite, Liu2025PersonaFlow}.
Zheng et al. proposed DiscipLink~\cite{Zheng2024DiscipLink} to help researchers explore and summarize potentially relevant literature across different disciplines. 
Liu et al. introduced Selenite~\cite{Liu2024Selenite}, which supports readers in retrieving and reasoning about unfamiliar domain knowledge during sensemaking, enabling synthesis across multiple papers. 
To foster cross-disciplinary ideation, Liu et al. developed PersonaFlow~\cite{Liu2025PersonaFlow}, which simulates experts from multiple domains to support research idea generation.

Previous research has demonstrated the benefits of integrating knowledge from various disciplines into the research process.
However, few studies have examined how interdisciplinary knowledge can enhance researchers’ critical paper reading and thinking skills.
This work addresses this gap by designing and evaluating a reading interface that incorporates AI agents and integrates insights from multiple research domains to stimulate critical-thinking skills.

\subsection{Challenges and opportunities in AI-assisted critical paper reading}
Existing studies have examined various AI-supported technologies that assist academic paper reading.
Most of them emphasize efficiency, helping researchers quickly extract key information~\cite{Chi2005ScentHighlights, Fok2023Scim, Fok2024Accelerating, Huth2024Eye, Gu2024An} and build connections between papers via inline citations~\cite{Zhang2008CiteSense, Rachatasumrit2022CiteRead, Chang2023CiteSee, Park2023QuickRef}.
However, relying on inline citations and history-based recommendations risks narrowing readers’ exposure, limiting opportunities to encounter papers from different disciplines and perspectives~\cite{Zheng2024DiscipLink}, and potentially reinforcing echo chamber effects~\cite{Chang2023CiteSee}, where people are repeatedly exposed to similar opinions.
Another key concern is over-reliance on AI, in which users increasingly delegate cognitive tasks to external systems, leading to cognitive offloading~\cite{carr2020shallows, sparrow2011google} and reduced active engagement in thinking. 
Over-reliance can negatively impact users’ independent thinking and learning processes~\cite{gerlich2025ai, lee2025impact}.
For example, Gerlich et al.~\cite{gerlich2025ai} demonstrated that reliance on AI tools undermines critical thinking performance: younger participants in their study exhibited higher dependence on AI and correspondingly lower critical thinking scores.
Yatani et al.~\cite{Yatani2024AIasExtraherics} argue that human–AI interactions should be designed to stimulate human higher-order thinking rather than encourage users to outsource cognition to AI systems.
These concerns highlight a challenge in research training: junior researchers, who typically have lower domain expertise, must carefully balance leveraging AI tools with cultivating independent critical judgment.

To promote critical thinking abilities, one of the most widely adopted critical reading strategies is the \reviseY{QRAC}  (\textbf{Q}uestion, \textbf{R}ead, \textbf{A}nswer, \textbf{C}heck) strategy~\cite{berkeley2013qrac,Keshav2007How, Peng2022CReBot}.
It encourages readers to ask questions of the text and seek answers that deepen understanding~\cite{wallace2021critical}.
Peng et al.~\cite{Peng2022CReBot} synthesized related self-questioning strategies and proposed a step-by-step guideline that supplies distinct question sets for each section of the paper reading process.
They found that the proposed system enhanced junior researchers’ engagement with paper reading.
Yuan et al. further improved this system by designing CriTrainer~\cite{Yuan2023CriTrainer}, which integrates questions dynamically generated by AI based on the contents of each section.
Their results suggest that the revised system helped students learn to formulate critical questions.
However, existing critical reading systems primarily engage readers through a limited set of viewpoints, which may constrain readers' exposure to alternative interpretations. Thus, to enable readers to be exposed to diverse viewpoints while reading a paper, we propose promoting thought exchanges between readers and AI agents.

Indeed, carefully-designed AI agents could introduce diverse viewpoints and foster users’ critical-thinking skills. 
For example, SeeWidely~\cite{SeeWidely2024} presents a multi-agent news reading interface where users interact with diverse social agents generated from news content and their respective personas, enabling multi-perspective engagement and mitigating filter bubbles.
ArgueTutor~\cite{Wambsganss2021ArgueTutor} enables students to practice argumentation skills through adaptive, dialogue-based tutoring.
Meanwhile, Tanprasert et al.~\cite{Tanprasert2024Debate} showed that debating with chatbots adopting different personas can affect users’ critical-thinking performance in online-video-platform contexts.
While prior work underscores AI’s double-edged role in supporting critical reading, it remains unclear how AI agents can actively scaffold users’ critical thinking throughout paper reading.
Different from efficiency driven LLM-augmented reading tools, our work positions AI agents as discussion partners that support junior researchers’ critical reading through structured thought exchanges, emphasizing reflective engagement over direct reliance on AI-generated outputs.


\section{Formative Study}

To explore potential interface designs for peer discussions during paper reading, we conducted a formative study with junior researchers.

\subsection{Participants}

We recruited eight participants \reviseY{(mean age = 25.5, SD = 2.1; 1 female, 7 male; P1--8)} using a snowball sampling approach through the authors’ social networks.
Participants were all students (3 master and 5 Ph.D. students; Appendix \ref{appendix: formative study}) from diverse disciplines. 
We compensated each participant with approximately 40 USD for two 60-minute studies. 
This formative study was approved by the Institutional Review Boards (IRB) of the first author's institution.

\subsection{\reviseY{Experimental Design and Procedure}}
\begin{figure*}[t]
  \centering
  \includegraphics[width=\linewidth]{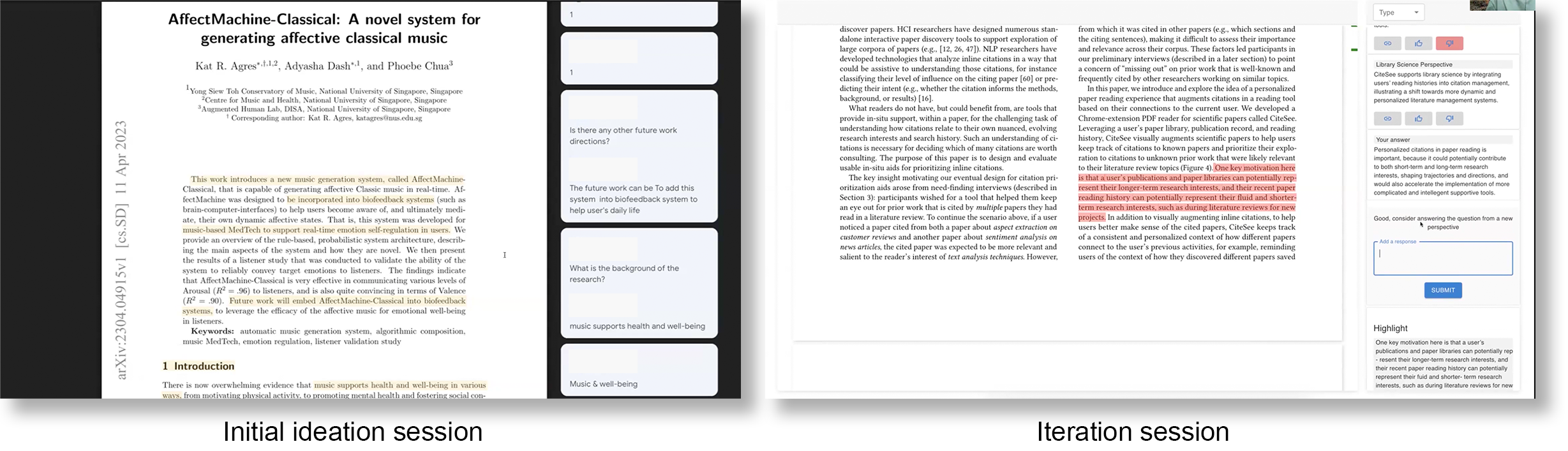}
  \caption{Formative study has two sessions: (1) left is \textit{initial ideation session}, participants used Google Doc to read the paper 
  with the experimenter to observe how participants interacted with both the PDF viewer and exchanged thoughts with reading peers; (2) right is \textit{iteration session}, based on the findings from the first session, we further implemented an initial prototype to gather more feedback.}
  \label{fig:formative_study} 
  \Description{An overview of the formative study setup with two sessions shown side by side.
The left panel illustrates the initial ideation session, where participants read an academic paper using a Google Docs-based PDF viewer while discussing and exchanging thoughts with the experimenter.
The right panel shows the iteration session, in which an early prototype was implemented based on insights from the first session and was used by participants to provide further feedback on the reading and discussion experience.}
\end{figure*}

The formative study consisted of two sessions: the \textit{initial ideation session} (Section \ref{sec: Intial ideation session}) and the \textit{iteration session} (Section \ref{sec: Refine ideation session}).
Each session consisted of three parts:
a 10-minute introduction, a 30-minute paper reading task, and a 20-minute interview.
The study was conducted online with Zoom.
For both sessions, we recorded and transcribed participants' utterances for qualitative analysis.

\subsubsection{Initial ideation session}
\label{sec: Intial ideation session}
The \textit{initial ideation session} aimed to understand how junior researchers use a standard digital paper reading interface and exchange thoughts with a peer.
We chose Google Docs as the probe (see Figure \ref{fig:formative_study}-left) because it offers features frequently used while reading scientific documents, such as highlighting and commenting~\cite{ReadingGroup99}.

We instructed participants to read a paper together with the experimenter. 
The experimenter acted as a reading peer by commenting on participants' annotations, proactively adding new ones, or raising questions.
This setting also allowed us to analyze their reading behaviors.
Participants were not constrained in how they interacted with Google Docs and were asked to think aloud throughout the reading process. 

Our findings showed that participants valued features such as raising section-based questions, offering cross-domain interpretations, and providing annotation feedback. These insights directly shaped our initial prototype, which implemented functionalities to scaffold critical reading and support multi-perspective understanding.

\subsubsection{Iteration session}
\label{sec: Refine ideation session}
We developed a low-fidelity prototype based on the feedback we gained from the previous session.
This low-fidelity prototype allowed participants to engage in peer reading with AI agents (see Figure \ref{fig:formative_study}-Right).
The design simulated a chat-based LLM interface that users could interact with while reading a paper in another window.
These agents could generate questions, answer questions, and interpret participants' highlights from the perspectives of different disciplines.
We asked participants to interact with the prototype and provide feedback on the format, timing, and nature of the AI agents’ responses.
Participants suggested that the agents’ answers should extend beyond the paper’s content to highlight domain-specific knowledge, and that the interface should include additional hints and guidance for junior researchers.

\subsection{Design Goals
}
By combining qualitative data from both sessions, we derived the following four design goals to support paper reading.

\subsubsection{\textbf{DG1}: Cross-Disciplinary Thought Exchange}

Four participants (P1, P3, P5, and P8) emphasized the importance of exchanging thoughts and opinions beyond the content of the paper to foster critical thinking skills: \textit{``For a beginner, it is necessary to look at it [the text] from different perspectives''} (P1).
Four (P3, P6, P7, and P8) suggested that generated AI agents' thoughts should be grounded in theories or domain-specific understandings:  
\textit{``If you mentioned some cognitive psychology stuff, you'd better give me very specific knowledge in cognitive psychology: its theories, results, and ways of thinking''} (P6).
However, three participants (P1, P6, and P8) suggested differently. They suggested the AI agents’ thoughts should remain relevant to the content of the paper, as it may otherwise \textit{``distrust''} (P8) readers.

\subsubsection{\textbf{DG2}: Section-Based Critical Thinking Guidance} 
Three participants (P1, P4, and P7) expressed the need for guidelines before starting to read: \textit{``When I first came in, I didn't even know how to read this article. At the beginning, it is necessary to establish a methodology for reading a paper''}  (P1).
Another participant noted that prompting questions before each section was effective, since sections contain relatively independent content, creating a \textit{``cognition gap''} (P3).
One junior researcher commented on this approach: \textit{``Asking a question is a great way to improve critical thinking''} (P7).

\subsubsection{\textbf{DG3}: Cooperative Support Without Disruption}
Balancing the timing and format of AI agents’ interventions is crucial to avoid disrupting the user’s reading flow.
Participants (P1, P2, P4, and P5) mentioned that interaction frequency should be based on the user context.
For example, participants suggested not generating questions during active reading behaviors, such as highlighting or commenting,  \textit{``There's no need to raise some questions and go into this''} (P5), because it could be distracting: \textit{``I may spend a lot of time staying here''} (P1).
However, participants also mentioned that interpretation from AI agents may help users' understanding: \textit{``If I do not understand part of the original sentence, then from another perspective, when it is reinterpreted, I may understand''} (P5).

\subsubsection{\textbf{DG4}: Follow-up Key Sentence Highlighting}
Furthermore, six participants (P1, P3, P4, P5, P7, and P8) wanted the system to provide hints, such as highlighting the important parts of the academic paper to assist paper reading:
\textit{``That’s a common issue for many junior researchers like me. I tend to dive straight into the paper and lose focus, not because of distractions, but because I lose track of the main structure of the paper''} (P3).
While such guidance could help users stay oriented, it also risks reducing their active engagement and critical thinking.
Therefore, features should be designed so that readers first read and think on their own, and only afterward receive prompts that remind them to \textit{``check if something important is missed''} (P5), or highlight additional insights.


\section{Interface Design}
\label{sec: Interface Design}
\begin{figure*}[t]
  \centering
  \includegraphics[width=\linewidth]{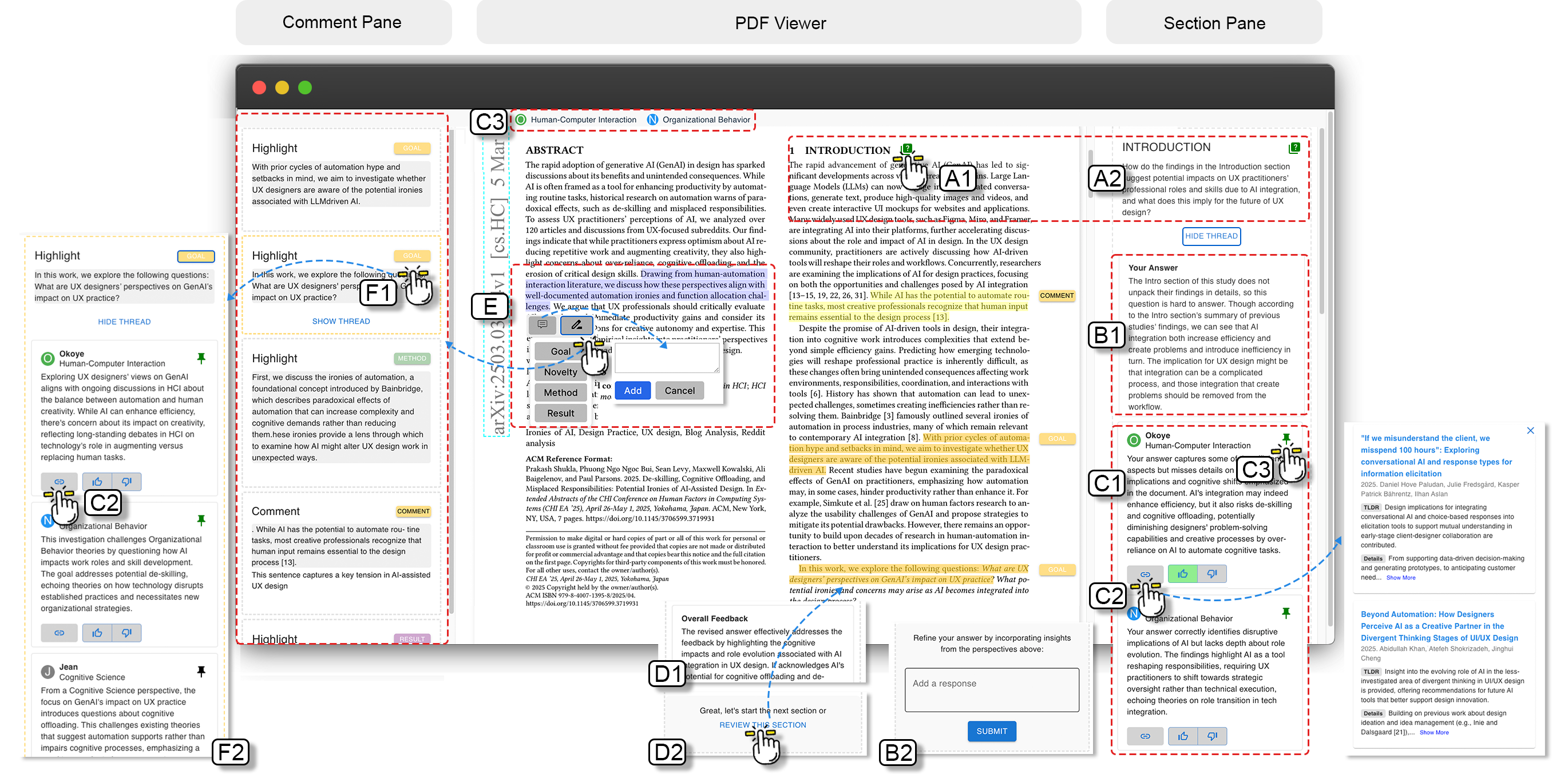}
  \caption{The user interface consists of a custom PDF viewer, a left \textit{Comment Pane}, and a right \textit{Section Pane}. Users can highlight sentences, leave comments, generate critical thinking questions, and engage in multi-turn thought exchanges with AI agents. }
  \label{fig:system_intro} 
  \Description{
  An overview of the system interface.
The center area shows a custom PDF viewer displaying an academic paper.
On the left, a Comment Pane lists user highlights and comments associated with specific sentences in the paper.
On the right, a Section Pane supports generating critical thinking questions and engaging in multi-turn thought exchanges with AI agents.
Users can highlight text in the PDF, leave comments, and interact with the agents in this interface.
}
\end{figure*}

We designed an interface (Figure~\ref{fig:system_intro}) that supports critical reading by enabling short question-and-answer exchanges with LLM-enabled AI agents.
Our interface consists of a custom PDF viewer for paper reading and two types of side panes, allowing the reader to engage in such exchanges in parallel with their reading.
The \textit{Comment Pane} (left pane) displays users' comments and highlighted text, followed by additional thoughts given by AI agents.
The \textit{Section Pane} (right pane) allows users to have a focused discussion about a selected section (``Introduction'' in the example of Figure~\ref{fig:system_intro}).
In this manner, our system supports two modes of thought exchanges during paper reading.
The following subsections describe the main functionalities of the two panes and how they implement the design goals identified earlier.

\subsection{Comment Pane}
\label{sec: Comment Pane}
Users can highlight a particular sentence in a color (\autoref{fig:system_intro}-E, Right button) and select one of the labels from the predefined set (Goal, Novelty, Method, and Result), also shown as cards in the \textit{Comment Pane}.
Users can also add comments after selecting a sentence (\autoref{fig:system_intro}-E, Left button), which will appear as a separate card labeled \textit{Comment}.

To reflect \textbf{DG1} and \textbf{DG3}, 
AI agents representing different research domains reinterpret the meaning of highlighted sentences or respond to user comments 
(\autoref{fig:system_intro}-F1, F2) only when users click the corresponding labels of the cards.
They can also add reactions and view references relevant to the corresponding comment (\autoref{fig:system_intro}-C2).

\label{sec: Interface Design}
\begin{figure*}[t]
  \centering
  \includegraphics[width=\linewidth]{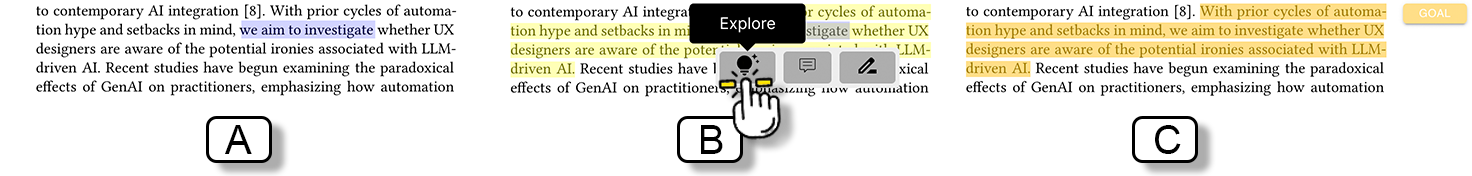}
  \caption{When users select a sentence (A) that overlaps with a pre-tagged sentence, the system automatically highlights the corresponding pre-tagged sentence (B). Users can then add the sentence to the \textit{Comment Pane} and maintain the highlight in the PDF viewer (C).}
  \label{fig:explore feature} 
  \Description{
  An example of the sentence selection and highlighting workflow.
Label A indicates a user-selected sentence in the PDF viewer.
When the selected sentence overlaps with a pre-tagged sentence, label B shows that the corresponding pre-tagged sentence is automatically highlighted by the system.
Label C illustrates how the selected sentence is added to the Comment Pane while the highlight is preserved in the PDF viewer.}
\end{figure*}

To reflect \textbf{DG4}, we pre-label sentences automatically highlighted by Semantic Reader \footnote{\url{https://www.semanticscholar.org/product/semantic-reader}}~\cite{Fok2023Scim} with Goal, Method, or Result, but these labels are hidden from the users.
When the users select a sentence that overlaps with a pre-tagged one (\autoref{fig:explore feature}-A), the corresponding pre-tagged sentence will automatically highlight (\autoref{fig:explore feature}-B), then the users can click the \textit{Explore} button and add this sentence to the \textit{Comment Pane} and leave the highlight in the PDF viewer (\autoref{fig:explore feature}-C).

\subsection{Section Pane}
\label{sec: Section Pane}
The \textit{Section Pane} allows users to interact with critical thinking questions. Users can trigger questions, submit their answers, and reflect on their responses within this pane. 
Following \textbf{DG2}, we adopt QRAC, a widely-used self-questioning method to enhance critical reading~\cite{berkeley2013qrac} as the main framework in the \textit{Section Pane}.
Before reading a section, users can click the question button (\autoref{fig:system_intro}-A1) next to the section title to generate critical thinking questions in the \textit{Section Pane} (\autoref{fig:system_intro}-A2).
Users are then prompted to enter their own answers.
To reflect DG1, AI agents subsequently assess the user’s answer and generate their own (\autoref{fig:system_intro}-C1).
Users can then compare these different answers, add reactions, and view related references, similar to the \textit{Comment Pane} (\autoref{fig:system_intro}-C2).
Users can pin a specific agent in the top bar, allowing that agent to continue the dialogue with the user in subsequent interactions (\autoref{fig:system_intro}-C3).
After reviewing the AI-generated answers, users are encouraged to revise their initial responses in the subsequent text entry field (\autoref{fig:system_intro}-B2).
The system then provides overall feedback on the revised answers (\autoref{fig:system_intro}-D1).

Following \textbf{DG4}, to help users better review the current section and fill in the information they may have missed, users can click the "REVIEW THIS SECTION" button ( \autoref{fig:system_intro}-D2) after receiving overall feedback.
This automatically highlights all pre-tagged sentences in the section, helping users review and identify information they may have missed.  

\subsection{Agent Design}
We designed the agent personas to represent distinct disciplinary perspectives relevant to the context of the current paper, scaffolding readers’ critical reading and thought exchanges with domain-specific knowledge. 
To balance users' cognitive load between reading and interacting with the agent, we designed the agents to provide responses only after users triggered events, and the agent would not engage in multi-turn conversations with the users. The prompts are in Appendix \ref{Appendix: Prompts}.

\subsection{System Implementation}
\label{sec: System Implementation}

\begin{table*}[t]
\centering
\caption{Examples of Critical Thinking Questions}
\label{tab:question_examples}
\begin{tabular}{|l|p{4cm}|p{6cm}|}
\hline
\textbf{Section} & \textbf{CReBot~\cite{Peng2022CReBot}} & \textbf{Ours} \\
\hline
Paper 14 Introduction~\cite{Adamo2024SoniWeight}  & How novel is the contribution? & How do the findings from the introduction suggest potential implications for future research directions or applications in the field of multisensory technologies?  \\
\hline
Paper 16 Discussion~\cite{Liu2024When} & Are these implications really insightful? & How might the design implications of using AI in family expressive arts therapy inform future technological developments in human-computer interaction beyond therapeutic contexts?  \\

\hline
\end{tabular}
\label{tab:critical_questions}
\end{table*}

\begin{table}[t]
\centering
\caption{Comparison of the quality of generated critical thinking questions with the baseline}
\begin{tabular}{lccc}
\toprule
Criterion & p-value & \multicolumn{2}{c}{Score Mean (SD)} \\
\cmidrule(lr){3-4}
{} & {} & Ours &  CReBot~\cite{Peng2022CReBot} \\
\midrule
Understandability & $<.001$  & 4.03 (0.40) & 4.44 (0.57) \\
Relevance         & $<.001$  & 4.29 (0.33) & 3.55 (0.43) \\
Criticalness      & $<.001$  & 4.11 (0.41) & 3.23 (0.49) \\
Comprehensiveness & $<.001$  & 4.06 (0.36) & 3.02 (0.47) \\
\bottomrule
\end{tabular}

\label{tab:critical_thinking_quesitons}
\end{table}

We implemented the system as a browser-based web application that supports both PDF rendering and interactive features. 
For the front-end, we used React.js \footnote{\url{https://react.dev/}} and TypeScript \footnote{\url{https://www.typescriptlang.org/}} for development, with Vite \footnote{\url{https://vite.dev/}} for project building, and the React-PDF-Viewer library \footnote{\url{https://react-pdf-viewer.dev/}} to render PDFs.
For the back-end, we used Python Flask to handle the server-side logic and API requests.
We used Grobid~\cite{GROBID} to parse academic papers' PDFs into structured XML files containing both content and layout information. These XML files were then provided to the OpenAI Assistants API (beta) \footnote{\url{https://platform.openai.com/docs/assistants/migration}}, which supports retrieval over uploaded documents. Through this setup, the system could generate questions and feedback grounded in the paper content.

A key implementation component is the generation of critical-thinking questions.
Previous work on critical reading training systems, such as CReBot~\cite{Peng2022CReBot}, analyzed and summarized self-generated questions into four categories: what, how, why, and how well.
CRiTrainer~\cite{Yuan2023CriTrainer} further divided these four categories into lower-order thinking questions (what and how), whose answers can be directly found in the text, and higher-order thinking questions (why and how well), which facilitate evaluation and creation —key aspects of critical thinking.
Following CRiTrainer's~\cite{Yuan2023CriTrainer} pipeline, we extracted 163 why and how well questions from the CReBot~\cite{Peng2022CReBot} database.
We used these examples for few-shot learning with GPT-4-turbo to generate section-specific critical-thinking questions.
The original and generated question examples are listed in Table \ref{tab:question_examples}.

To evaluate the quality of the generated questions, we adopted the criteria of understandability, relevance, criticalness, and comprehensiveness, following prior work~\cite{le2015evaluation, Peng2022CReBot}. We randomly selected 16 papers from different sessions of a top-tier HCI conference \footnote{ACM CHI 2024 \url{https://chi2024.acm.org/}}.
For each section of these papers, we generated critical thinking questions, resulting in a total of 123 generated questions.
Twelve senior Ph.D. researchers evaluated each generated question against the original human-labeled dataset using a 5-Likert scale on the four criteria.
We applied the Wilcoxon signed-rank test to compare the two sets.

The generated critical thinking questions outperformed the original human-labeled dataset in three dimensions: relevance, criticalness, and comprehensiveness (shown in Table \ref{tab:critical_thinking_quesitons}).
However, the understandability of the generated questions was lower than that of the baseline. 
This could be due to their higher information density and more complex sentence structures, which increase cognitive load and make them less immediately readable compared to human-written questions.
\revise{Thus, it might be a trade-off between depth of content and readability.}


\section{User Study}

To examine how reading papers with single-agent and multi-agent thought exchanges affects readers’ critical thinking (\textbf{RQ1}) and critical-reading practices (\textbf{RQ2}), we conducted a between-subjects experiment consisting of a pre-training test, a two-week training phase, and a post-training test.
We randomly assigned our participants to one of three conditions:
(1) thought exchanges with a single disciplinary agent (\condS);
(2) thought exchanges with multiple disciplinary agents (\condM);
and (3) no interface for thought exchanges (\condN), where participants read and answered questions only with static feedback.
The two-week duration allowed us to observe changes in participants’ reading practices and critical-thinking abilities over time.
We obtained approval on the user study described in this section from the IRB of the first author’s institution.

\begin{figure*}[t]
  \centering
  \includegraphics[width=\linewidth]{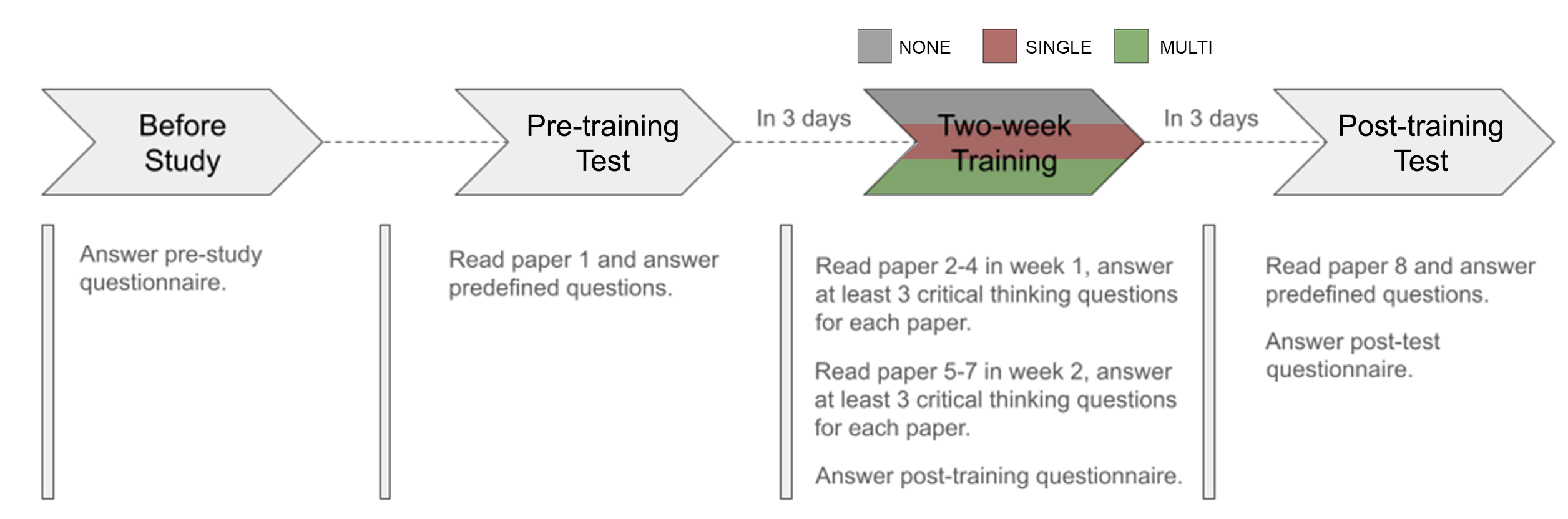}
  \caption{User study procedure}
    \label{fig: user study flow}
  \Description{The User study procedure consists of four parts: (1) participants need to answer pre-study questionnaire to register the user study; (2) then participants need to complete pre-training test to read 1 paper without interface support; (3) after that, participants need to complete two weeks training periods to read another 6 papers with the interface support; (4)finally, participants need to complete post-training test to read the last paper without the interface support.}
\end{figure*}

\subsection{\reviseY{Experimental Design}}
Participants were instructed to read the assigned papers and answer critical thinking questions under one of the three conditions.
To support and measure critical engagement, participants responded to critical thinking questions, which were either pre-generated and assigned by experimenters (during the pre-training and post-training phases) or system-generated (during the training phases).

\textit{Multiple-agent Condition (\conditionM)}:
Previous research has suggested that interdisciplinary knowledge can enhance researchers’ performance~\cite{van2015interdisciplinary} in tasks such as literature searching~\cite{Zheng2024DiscipLink} and idea generation~\cite{Liu2025PersonaFlow}.
To emulate thought exchanges with researchers in different fields, we employed LLM prompts to design AI agents representing different research domains.
In this condition, LLM dynamically chose three research domains based on the content of a section. 
Once a user pinned (\autoref{fig:system_intro}-C3) an agent, it provided feedback consistently from that disciplinary perspective throughout the interaction.
The prompts used in our implementation are in Appendix \ref{Appendix: Prompts}.

\textit{Single-agent Condition (\conditionS)}:
Participants interacted with one AI agent representing a researcher from a fixed domain distinct from their own discipline.
We included this condition because this 1-on-1 setup is a typical format in existing AI or LLM services, 
and it also allows us to examine whether involving multiple agents provides additional benefits.
We removed interface components for comparing or evaluating thoughts by multiple AI agents, such as the \textit{pin} (\autoref{fig:system_intro}-C3), \textit{thumb up}, and \textit{thumb down} (\autoref{fig:system_intro}-C2, Middle and Right) functions.

\textit{No Agent Condition (\conditionN)}:
This condition served as a reference where participants would not interact with any AI agent.
Instead, participants received a static text-based reminder (shown in Appendix \ref{appendix: static feedback}) encouraging them to revise and improve their initial answers, following the QRAC approach~\cite{berkeley2013qrac}.
No reaction features (\autoref{fig:system_intro}-C2, C3) were available in this condition.

\subsection{Participants}
We recruited 46 students (mean age = 23.35, SD = 3.74; 11 male, 33 female, 2 undisclosed; 27 master’s students, 19 undergraduates) who majored in or showed interest in Human-Computer Interaction (HCI) and related fields such as computer science, sociology, or psychology.
We chose HCI because its interdisciplinary nature makes it especially likely to benefit from multiple perspectives.
Fifteen participants were assigned to each of the multi- and single-agent conditions, and sixteen to the no-agent condition.

We chose participants who had English proficiency equivalent to a TOEIC score of 700 or above, ensuring sufficient ability to comprehend academic papers written in English.
Additionally, participants need to either self-report having no prior critical reading experience to the recruiting questionnaire question \textit{``Have you received any training in critical reading?''} or, if they did have such experience, to self-rate their critical reading ability as average or below to the question \textit{``If yes, how would you rate your critical reading skills?''}.
Participants were recruited through a combination of the authors’ social networks, social media platforms, and dedicated participant recruitment services to ensure a diverse and sufficient sample.
Each participant received compensation equivalent to approximately 65 USD for participating in the study.

\subsection{Procedure}

The user study consisted of three phases (shown as Figure \ref{fig: user study flow}):
a 1.5-hour pre-training test (Day 1), a two-week critical reading training period (Day 2–15), and a 1.5-hour post-training test (Day 16).
The pre- and post-training tests were conducted within three days before and after the training period, respectively.
To mitigate order effects, papers 1 and 8 were counterbalanced so that half of the participants received them in reverse order.

The interface was deployed online using a Google Cloud Platform (GCP) Compute Engine\footnote{\url{https://cloud.google.com/compute}}, and participants conducted the experiments on their own personal computers.
To ensure that participants did not use external generative AI tools during reading, we informed them that AI usage traces in their answers would be monitored and asked them to record their screens and submit recordings after the study.
Aside from the designated reading interface, participants were only permitted to access digital libraries such as Google Scholar\footnote{\url{https://scholar.google.com}}, as well as translation tools like Google Translate\footnote{\url{https://translate.google.com}}.
Eight CHI Late-Breaking Work papers (Appendix \ref{ref:appendix_LBW}) were randomly selected as reading materials.
All papers addressed topics related to large language models (LLMs), AI, or chatbots. 
These papers were relatively short and accessible, making them appropriate for junior researchers from diverse disciplinary backgrounds.
In consideration of ethical concerns surrounding the upload of potentially copyrighted PDFs to cloud-based LLMs, we limited our user study materials to publicly available arXiv \footnote{\url{https://arxiv.org/}} versions.

\subsubsection{Pre-training test.}  
We designed and conducted a pre-training test to evaluate participants’ critical thinking ability before using our system.
A CHI Late-Breaking Work paper was designated as the reading material during this phase.
Participants received a list containing one critical-thinking question per paper section and provided written responses (see examples of questions in Table \ref{tab:question_examples}).
Participants first reviewed each question, read the corresponding section, and then answered it.
The entire pre-training test was completed within 1.5 hours.

\subsubsection{Two-week critical reading training.} 
Before the training phase began, each participant was provided with a tutorial page and instructed to read it carefully.
Participants were then randomly assigned to one of three experimental conditions: the \condM, the \condS, or the \conditionN.
Then, participants began the two-week session during which they were instructed to critically read three designated CHI Late-Breaking Work papers per week (for a total of 6 over 2 weeks). 
Their task was to generate at least three critical-thinking questions from random sections, answer them, and then revise their initial answers based on feedback.
In total, each participant read six papers and answered at least 18 critical thinking questions throughout the training period.
Participants' answers and interaction behaviors were logged for subsequent analysis.
Participants were permitted to use only the designated interface and, if needed, digital libraries and translation tools as in the pre-training test.
Participants were free to use any of the interaction features embedded in the system. 
Answering critical thinking questions was a mandatory task, while other active reading features (e.g., highlighting, commenting) were optional and could be used according to individual reading habits.
After completing all six papers, participants were asked to complete a questionnaire to provide feedback on:
(1) their self-assessed critical thinking ability, and
(2) their perceived cognitive load when using different system conditions.

\subsubsection{Post-training test}
\label{sec: post-training test}
To evaluate the learning effects of our interface, the post-training test was conducted within three days after the completion of the training phase.
In this session, participants were given 1.5 hours to read a new CHI Late-Breaking Work paper and answer a set of predefined critical thinking questions in the same flow as in the pre-training test.
Additionally, participants were asked to respond to open-ended questions reflecting on their experience with the system: (1) \textit{``Which feature in this system was the most helpful for your critical reading of papers? Please explain why''}, (2) \textit{``Which skills do you think have improved after these two weeks of training? Please describe how they have improved''}, (3) \textit{``Which feature did you find the least helpful or least satisfying? Please explain why and how it could be improved''}.

\subsection{Measurement}

\subsubsection{Response Quality in Pre and Post-training Test (RQ1)}
Four Ph.D. students specializing in HCI, 
who had previously evaluated the quality of the generated critical thinking questions (Section \ref{sec: System Implementation}), assessed the quality of participants’ answers to critical thinking questions in the pre-training test and post-training test.
They evaluated participants' responses to each question in the papers based on the Critical Thinking Value Rubric~\cite{AACU2009_CriticalThinking}, a standardized assessment tool for evaluating critical thinking skills across five dimensions: (1) Explanation of Issues, (2) Use of Evidence, (3) Influence of Context and Assumptions, (4) Student’s Position (Perspective, Thesis/Hypothesis), and (5) Conclusions and Related Outcomes.
Each dimension is scored on a 4-point scale (1 to 4), with higher scores indicating more advanced critical thinking ability. 
\reviseY{For analysis, the final score for each dimension was computed as the average of the scores given by the four reviewers.
The total critical thinking score was then calculated as the average of these dimension-level average scores.
}
The order of responses was randomized so that evaluators could not identify whether an answer came from the pre- or post-training test.

\subsubsection{Number of Unique Viewpoints (RQ1)}
\label{sec: unique viewpoints}
As discussed in Section \ref{sec: related work 1}, considering multiple perspectives is crucial for critical thinking~\cite{dekker2020teaching}.
Therefore, to address RQ1, we analyzed the number of unique viewpoints expressed in participants' textual answers during the two-week training period, using sentence-level semantic clustering. 
Following previous work that classified and grouped similar concepts~\cite{pasch2025human},
we segmented participants' answers into sentences using a standard sentence tokenizer~\cite{bird2004nltk}.
Then, each sentence was encoded into a dense vector representation using a pretrained sentence embedding model: SentenceTransformer~\cite{reimers2019sentence} with the 'paraphrase-MiniLM-L6-v2' model.
To estimate the number of unique viewpoints, we applied K-Means~\cite{pedregosa2011scikit} clustering on the sentence embeddings. 
The number of clusters was determined automatically by the point where the second-order difference of the inertia values (as cluster number increases) is largest, which corresponds to the “elbow”~\cite{kodinariya2013review} in the curve. 
The resulting cluster count was used as an indicator of the number of unique viewpoints within the response. 
We aggregated these counts per participant and computed the average number of viewpoints for both initial answers and improved answers during the training phase.

\subsubsection{Perceived Critical Thinking (RQ1)}
We measured participants' level of critical thinking via The Critical Thinking Self-Assessment Scale (CTSAS)~\cite{Payan2022Development}, a widely used self-report instrument rated on a 7-point Likert scale (1 to 7: 1 represents strongly disagree, 7 represents strongly agree).
Following the approach of Tanprasert et al.~\cite{Tanprasert2024Debate}, who extracted high-demand items to simplify the original CTSAS (n = 12), we also selected 12 relevant items from the original CTSAS that were most appropriate for our system context (\reviseY{see Appendix  \ref{ref:appendixA}}). 
The CTSAS scale consists of six dimensions: \textit{Interpretation}, \textit{Analysis}, \textit{Evaluation}, \textit{Inference}, \textit{Explanation}, and \textit{Self-regulation}, each represented by two items. 
For analysis, we averaged the scores of the two items to compute the score for the corresponding dimension.
Following previous work~\cite{Tanprasert2024Debate}, we derived the total critical thinking score by averaging the scores of the six dimensions.
Participants completed this self-assessment questionnaire after the training phase to reflect on their perceived critical thinking ability.

\subsubsection{Perceived Workload (RQ1)}

We further asked participants to report the perceived workload they experienced in the different \textsc{conditions}. 
We used the NASA-TLX~\cite{hart1988development} to allow participants to self-measure their workload.
\reviseY{For analysis, we computed the average score across the six NASA-TLX subscales, following common practice~\cite{hart1988development}.}

\begin{figure*}[t]
  \centering
  \includegraphics[width=.8\linewidth]{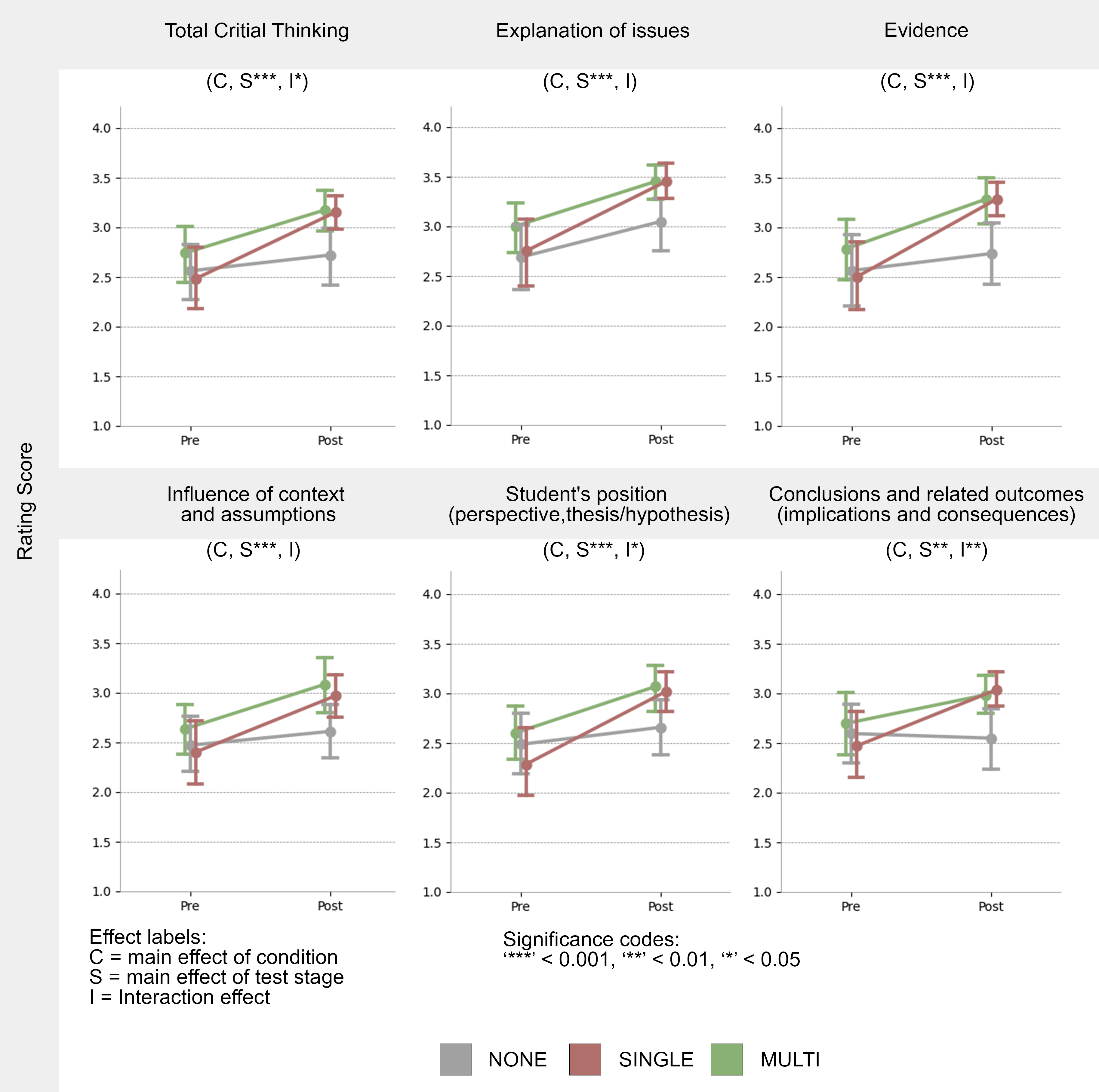}
  \caption{External review scores in five dimensions of the Critical Thinking Value Rubric, and their average as total critical thinking scores, across three groups and pre-/post-training tests.}
  \label{fig:external_score} 
  \Description{A quantitative comparison of external review scores using the Critical Thinking Value Rubric.
  Scores for five dimensions and an overall average are shown across three participant groups, with measurements taken before and after the training. The results indicate two-week training under the multi and single condition effectively enhanced critical thinking ability, together with no significant change in the none condition.}
\end{figure*}

\subsubsection{Interaction Behavior (RQ2)}
\label{sec:interaction behavior}
To explore differences in interaction behavior between the \conditionS and \conditionM, we analyzed how participants revised their answers from initial to revised responses to examine differences in reflective behavior. 
We calculated the average time participants took to improve their answers based on the feedback, as well as how they interacted with the system. 
We categorized highlights, comments (\autoref{fig:system_intro}-E), automatically generated highlights (\autoref{fig:explore feature}), and user-triggered AI agent comments from the \textit{Comment Pane} (\autoref{fig:system_intro}-F1) as \textit{Annotations}, as all of them require participants to actively engage with the paper content. 
Similarly, we categorized checking references, thumbs up, thumbs down (\autoref{fig:system_intro}-C2), and pinning agent (\autoref{fig:system_intro}-C3) as \textit{Reactions}, as they all require participants to evaluate and analyze the statements from different disciplinary perspectives. \revise{Here, checking references was considered a \textit{Reaction} in the sense that participants used it to critically evaluate or verify the credibility of the statements within the card components.} 
Inspired by the previous work on interaction behavior analysis~\cite{Alvina2020Where}, we then mapped each participant’s behavior data onto a scatter chart to visualize their \textit{Annotation} and \textit{Reaction} behaviors.

\subsubsection{Open-Ended Question Analysis (RQ2)}
We further conducted a thematic analysis~\cite{braun2006using} of the three open-ended questions (listed in Section \ref{sec: post-training test}) in the post-training test questionnaire. 
For Question 1, we employed a combined top-down and bottom-up approach.
Specifically, the top-down analysis mapped participants’ responses onto the six established dimensions of cognitive thinking defined in Bloom's Taxonomy~\cite{bloom1956taxonomy}. In contrast, the bottom-up analysis enabled us to identify emergent themes beyond the predefined framework.
For Questions 2 and 3, we adopted a purely bottom-up approach, as these questions were exploratory in nature and aimed to capture participants’ reflections and experiences without imposing prior categorizations.


\section{Results}

\begin{figure*}[t]
  \centering
  \includegraphics[width=.6\linewidth]{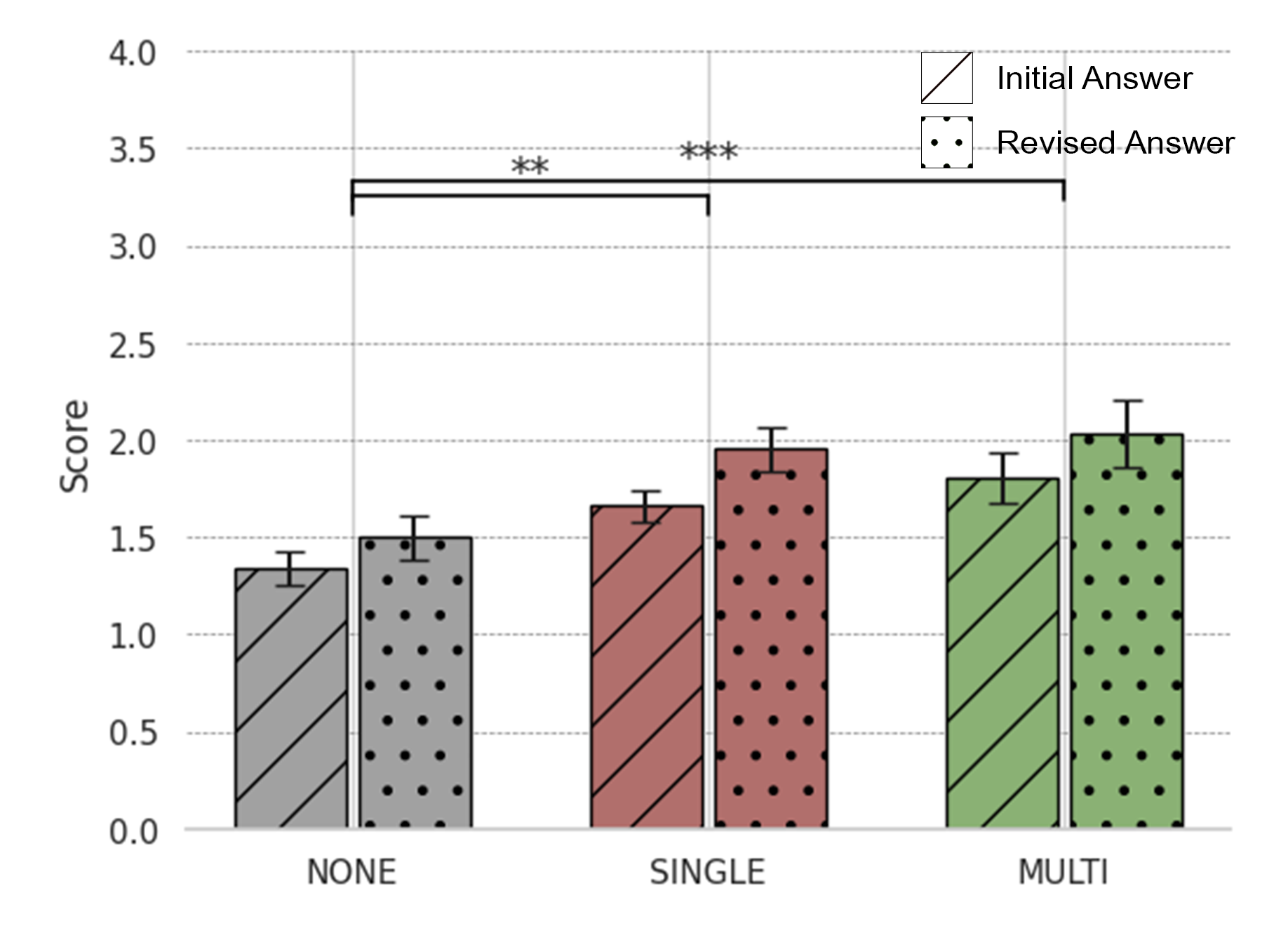}
  \caption{Average number of K-means clusters for each group at each time point (Initial vs Revised). Significance levels are indicated by asterisks: * $p < .05$, ** $p < .01$, *** $p < .001$.}
  \label{fig:viewpoints} 
  \Description{
  A quantitative comparison of the average number of clusters across participant groups.
  Values are shown for two time points, Initial and Revised, with statistical significance marked using asterisks. The results indicate that participants in single and multi condition generated more unique viewpoints than those in none condition, while there was no significant difference between single and multi condition.
  }
\end{figure*}

\begin{figure*}[t]
  \centering
  \includegraphics[width=\linewidth]{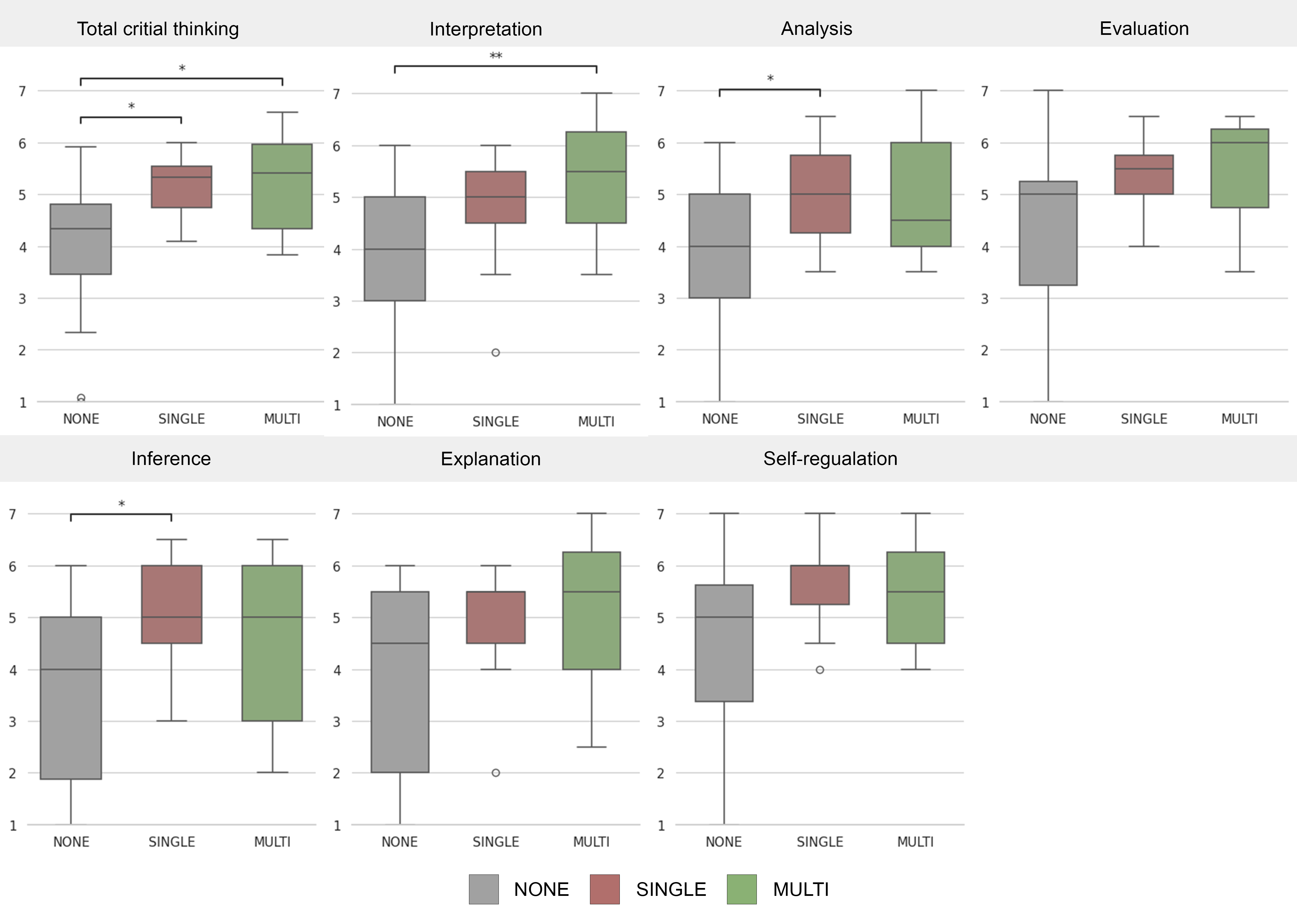}
  \caption{Participants' self-critical thinking assessment scores. Significance levels are indicated by asterisks: * $p < .05$, ** $p < .01$.}
  \label{fig:self_assessment}
  \Description{A quantitative comparison of participants' self-assessed critical thinking scores.
  Results are presented across different conditions, with statistical significance marked by asterisks. Overall, Participants in both the single and multi condition reported higher self-perceived critical thinking ability than those in the none condition, with significant improvements observed across total scores as well as the Interpretation, Analysis, and Inference dimensions}
\end{figure*}

\subsection{Response Quality Using Critical Thinking Score in Pre and Post-training Test (RQ1)}
\label{sec: external review}

We conducted a 3 (\condN, \condS, \condM; between-subject) × 2 (pre vs. post; within-subject) aligned rank transform mixed-design ANOVA on the average pre-training and post-training test scores evaluated by external reviewers. 
The five-dimensional critical thinking scores and total critical thinking scores across pre- and post-training tests are illustrated in Figure \ref{fig:external_score}. 
The full results are in Appendix \ref{Appendix: User Study Result}. 
Prior to the statistical analyses, we assessed inter-rater reliability among the four external reviewers using intraclass correlation coefficient (ICC), specifically ICC(3,k)~\cite{shrout1979intraclass} for five dimensions. 
The resulting ICC values indicated moderate to good agreement across the five dimensions, ranging from 0.74 to 0.87 for the pre-training test and from 0.69 to 0.85 for the post-training test.

There was a significant main effect between the pre-training and post-training phases on total critical thinking scores.
There was also a significant interaction effect of the \textit{condition} and \textit{test stage}.
A significant improvement from the pre-training test to the post-training test stage was observed in the \conditionM ($t$(14) = 6.16, $p < .001$, Cohen's $d = 1.28$) and \conditionS ($t$(14)=4.65, $p<.001$, Cohen's $d=1.18$), but not in the \conditionN, according to the post-hoc ART-based pairwise comparisons.
Post-hoc analysis at the pre-training test showed that the three types of thought exchange conditions did not have significant differences in the critical thinking score. 

Our analysis revealed significant main effects of test stage across all five dimensions (see Figure \ref{fig:external_score}). 
Post-hoc analyses revealed that the \condM and \conditionS improved significantly on all dimensions. 
\conditionN showed a significant improvement only in the \textit{Explanation of Issues} dimension ($t$(15) = 2.997, $p < .01$, Cohen's $d = 0.67$), whereas no significant changes were observed for the other dimensions.
At the pre-training test stage, no significant differences were observed between the three conditions on any dimension, indicating comparable baseline levels. 

These results indicate two-week training under the \condM and \conditionS effectively enhanced critical thinking ability, together with no significant change in the \conditionN.

\begin{table*}[t]
\centering
\caption{Participants' NASA-TLX Scores Across Conditions and Kruskal–Wallis statistical effects}
\label{tab:mental_demand}
\begin{tabular}{lcccc}
\toprule
Criteria & Stat. & \multicolumn{3}{c}{Mean score (± SD)} \\ 
\cmidrule(lr){3-5}  
{} & {} & \hlcell[100]{np}{\textsc{none}} & \hlcell[100]{sp}{\textsc{single}} & \hlcell[100]{mp}{\textsc{multi}} \\
\midrule
Mental Demand  & $H=0.39$, $p=.84$ & 7.88 (1.67) & 7.80 (1.01) & 7.80 (1.26) \\
\midrule
Physical Demand  & $H=0.29$, $p=.86$ & 5.56 (2.63) & 5.13 (2.42) & 4.93 (2.43) \\
\midrule
Temporal Demand  & $H=0.07$, $p=.96$ & 4.81 (2.29) & 5.00 (1.89) & 5.07 (2.05) \\
\midrule
Performance  & $H=0.08$, $p=.96$ & 5.19 (2.51) & 5.33 (1.54) & 5.13 (2.26) \\
\midrule
Effort  & $H=0.80$, $p=.67$ & 7.38 (1.67) & 6.93 (1.39) & 7.27 (1.33) \\
\midrule
Frustration & $H=4.90$, $p=.09$ & 5.75 (2.65) & 4.67 (1.95) & 6.40 (2.44) \\
\bottomrule
\end{tabular}
\end{table*}

\subsection{The number of unique viewpoints (RQ1)}
\label{sec: viewpoints}

We collected a total of 880 valid pairs of initial and revised
answers from the two-week training phase (323, 277, and 280 pairs from \condN, \condS, and \conditionM, respectively).
To evaluate the diversity of viewpoints reflected in participants’ responses,
we conducted a $3$ (\condN, \condS, \condM; between-subject) $\times$ $2$ (Initial answer vs. Revised answer; within-subject) multifactorial mixed-design ANOVA on each participant’s average number of clusters obtained through the process described in Section \ref{sec: unique viewpoints}. 
The results (illustrated in Figure \ref{fig:viewpoints}) revealed a significant main effect of the interface conditions
($F[2, 43] = 5.68$, $p<.01$, $\eta_p^2 = 0.21$), and a significant main effect of time ($F[1, 43] = 23.66$, $p<.001$, $\eta_p^2 = 0.35$).

Post-hoc Tukey HSD test showed that the \conditionM had a significantly higher average number of clusters than the \conditionN ($\text{mean difference} = 0.50, SE = 0.12, T = 4.19, p < .001,$ Hedges’ $g = 0.98$).
Similarly, the \conditionS also had a significantly higher number of clusters than the \conditionN ($\text{mean difference} = -0.39, SE = 0.12, T = -3.26, p < .01,$ Hedges’ $g = -0.96$).
In contrast, we did not observe a significant difference between the \condM and \conditionS. 

The results indicate that participants in \condS and \conditionM generated more unique viewpoints than those in \conditionN, while there was no significant difference between \condS and \conditionM.

\begin{figure*}[t]
  \centering
  \includegraphics[width=.6\linewidth]{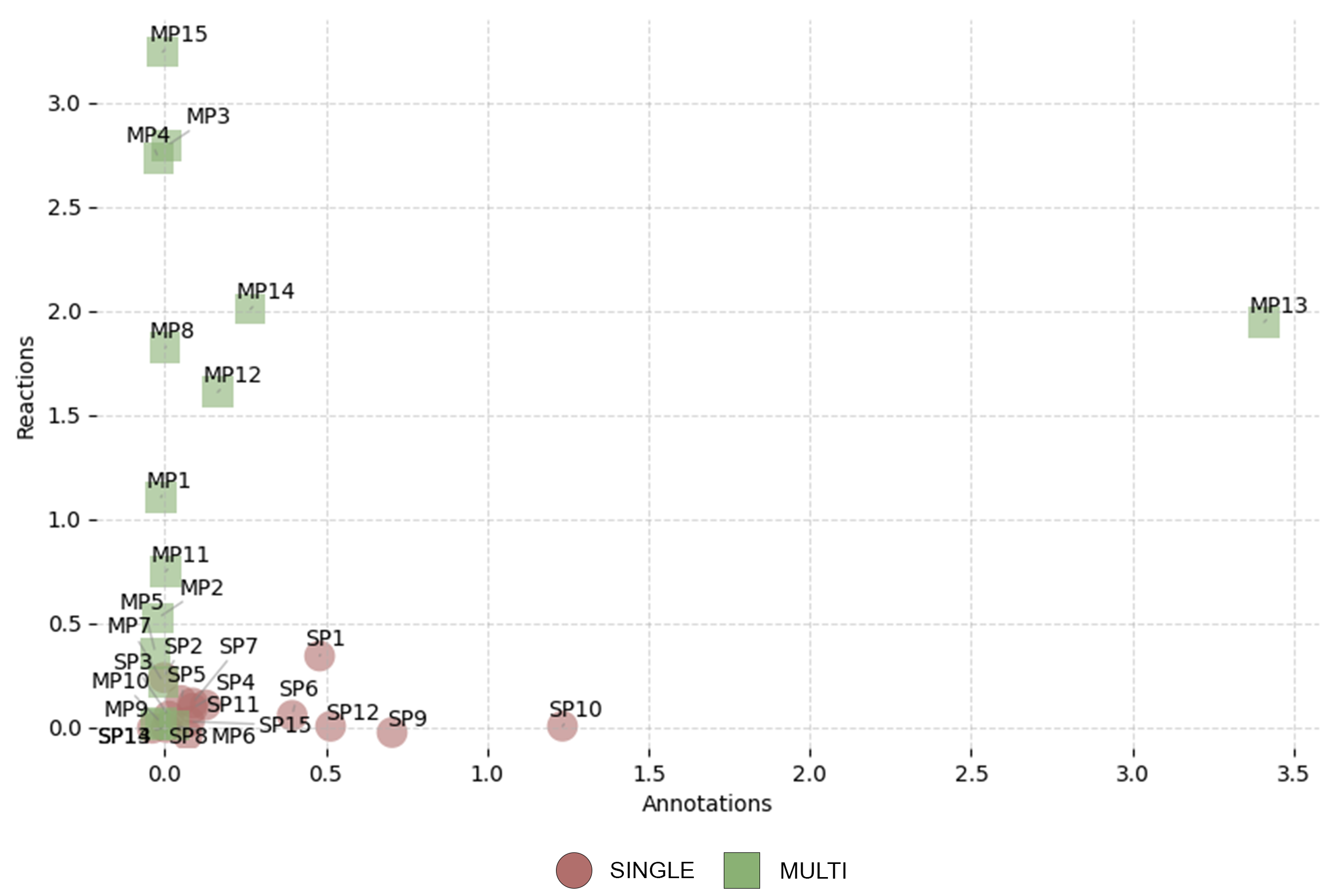}
  \caption{Scatter plot of participants’ interaction behaviors. The vertical axis represents the average usage counts of Reactions, and the horizontal axis represents the average usage counts of Annotations. The figure illustrates that participants in the \conditionM condition relied more on reactions and less on annotations 12 out of 15 participants in \conditionM never used annotations; while participants in the \conditionS engaged with annotations more frequently. 
  }
  \label{fig:interaction_log1}
  \Description{
  A scatter plot visualizing participants' interaction behaviors.
  Each point represents one participant.
  The horizontal axis shows the average number of annotation uses, and the vertical axis shows the average number of reaction uses.
  Participants from different experimental conditions are distinguished using different markers or colors.
  }
\end{figure*}

\subsection{Perceived Critical Thinking (RQ1)}
\label{sec: self critical thinking}

To examine the effects of different conditions on participants’ self-perceived critical thinking ability, we conducted a Kruskal–Wallis test for average score with conditions as a factor.
The results (see Figure \ref{fig:self_assessment}) indicated a significant difference between three conditions on the total critical thinking score ($\chi^2(2) = 8.37, p < .05, \eta^2 = 0.15$). 
Post-hoc pairwise comparisons using Dunn’s test with Holm correction showed that the \conditionN ($M=3.94, SD=1.44$) scored significantly lower than both the \conditionM ($M=5.16, SD=0.97, p < .05$) and the \conditionS ($M=5.13, SD=0.63, p < .05$).

We applied the same analysis to each subcomponent of critical thinking. 
For \textit{Interpretation},
the Kruskal-Wallis test found a main effect ($\chi^2(2) = 9.61, p < .01, \eta^2 = 0.18$) with post-hoc tests revealing significant differences ($p < .01$) between \condM ($M=5.40, SD=1.04$) and \conditionN ($M=3.88, SD=1.45$).
For \textit{Analysis}, the test found a main effect ($\chi^2(2) = 6.79, p < .05, \eta^2 = 0.11$) with post-hocs revealing significant differences ($p < .05$) between \condS ($M=5.03, SD=0.95$) and \conditionN ($M=3.78, SD=1.49$).
For \textit{Inference}, the test found a main effect of condition ($\chi^2(2) = 7.64, p < .05, \eta^2 = 0.13$), the post-hocs revealed that a significant difference ($p < .05$). between \condS ($M=5.10, SD=0.99$)  and \conditionN ($M=3.53, SD=1.72$).

Overall, Participants in both the \condS and \conditionM reported higher self-perceived critical thinking ability than those in the \conditionN, with significant improvements observed across total scores as well as the \textit{Interpretation}, \textit{Analysis}, and \textit{Inference} dimensions.

\begin{table*}[htp]
\centering
\caption{Summary of Themes, Examples, and Participants. 
Darker cells correspond to higher percentages. '-' indicates the feature is not available in the interface.}
\label{tab:themes}
\begin{tabular}{p{2.8cm} p{7.8cm} c c c} 
\hline
Themes & Examples & \#\condM & \#\condS & \#\condN \\
\hline
\multicolumn{5}{l}{\textbf{Which skills do you think have improved after these two weeks of training?}} \\
Reading Speed & \textit{``It is obvious that I can read the paper more focused, and the time of reading a paper is decreased.''} (NP2) 
& \hlcell[25]{mp}{2/15} 
& \hlcell[25]{sp}{3/15} 
& \hlcell[50]{np}{5/16} \\
\cline{2-5}
Concentration & \textit{``My reading speed has improved a lot, mainly because I can concentrate faster and better while reading now.''} (NP4) 
& \hlcell[25]{mp}{1/15} 
& \hlcell[25]{sp}{1/15} 
& \hlcell[25]{np}{3/16} \\
\cline{2-5}
Knowledge & \textit{``My knowledge has expanded after reading so many HCI papers.''} (NP9) 
& \hlcell[25]{mp}{1/15} 
& \hlcell[25]{sp}{2/15}
& \hlcell[25]{np}{3/16} \\
\cline{2-5} 
Comprehension & \textit{``I think the skill to summarize the whole paper has improved. With the help of AI feedback, I can summarize the paper as a whole, with a holistic view and a focus on more details.''} (SP4) 
& \hlcell[25]{mp}{1/15}  
& \hlcell[50]{sp}{6/15} 
& \hlcell[50]{np}{6/16} \\
\cline{2-5}
Application & \textit{``The skill is repeated reading... At first, I had no idea how to start, but after a few days, I learned to re-read the paper to gain more information.''} (NP11) 
& \hlcell[25]{mp}{1/15} 
& \hlcell[25]{sp}{1/15} 
& \hlcell[25]{np}{3/16} \\
\cline{2-5}
Analysis & \textit{``I have learned to analyse information in a multi-disciplinary lens. When I read, I will think about how the article is discussing theories or addressing problems in an HCI sense, or in a Social Science/Computer Science way.''} (MP12) & \hlcell[75]{mp}{8/15} 
& \hlcell[25]{sp}{2/15} 
& \hlcell[25]{np}{1/16} \\
\cline{2-5} 
Synthesis & \textit{``The skill to extract core information and reorganize it...''} (MP1) 
& \hlcell[25]{mp}{3/15}
& \hlcell[25]{sp}{2/15} 
& 0/16 \\
\cline{2-5}
Evaluation & \textit{``I am being confident in judging the relevance and usefulness of information in different research contexts.''} (SP1)  
& \hlcell[50]{mp}{6/15} 
& \hlcell[50]{sp}{5/15} 
& \hlcell[25]{np}{1/16}  \\
\hline

\multicolumn{5}{l}{\textbf{Which feature in this system was the most helpful for your critical reading of papers?}} \\
External Disciplinary Perspectives & \textit{``The system worked best when it reached the ideal of giving me an interesting new perspective which was wholly relevant to the angle at which I chose to answer the initial AI prompt.''} (MP2) & \hlcell[75]{mp}{11/15} & \hlcell[100]{sp}{12/15} & - \\
\cline{2-5}
Self Reflection & \textit{``After I answer a question, some questions will appear to make me rethink my answer.''} (NP14) 
& \hlcell[0]{mp}{0/15} 
& 0/15 
& \hlcell[50]{np}{4/16}  \\
\cline{2-5}
Q\&A & \textit{``The question is the most helpful part. It can help me understand the structure of each section.''} (NP9) 
& \hlcell[25]{mp}{2/15} 
& \hlcell[25]{sp}{1/15} 
& \hlcell[25]{np}{3/16}  \\
\cline{2-5}
Highlight & \textit{``Highlighting with a different color. It can help me classify different types of information and reorganize them.''} (NP16) 
& \hlcell[25]{mp}{2/15}
& \hlcell[25]{sp}{1/15} 
& \hlcell[50]{np}{5/16}  \\
\hline

\multicolumn{5}{l}{\textbf{Which feature did you find the least helpful or least satisfying?}} \\
Nitpicking Feedback & \textit{``All feedback is given in a ‘Yes, but…’ format, even when I feel that I have already addressed the aspect in sufficient detail... I lose confidence in providing a complete answer in a single attempt.''} (MP3) 
& \hlcell[50]{mp}{4/15} 
& 0/15 
& -  \\
\cline{2-5}
Contextual Drift & \textit{``some of the experts' comments contained information not included in the text, and some of the content was not necessarily directly related to the interpretation of the text.''} (MP13) 
& \hlcell[25]{mp}{3/15} 
& \hlcell[25]{sp}{1/15} 
& -  \\
\cline{2-5}
Static Feedback & \textit{``This system does not seem to have made any further evaluation of the given answers. It would be better if further evaluation could be provided or if the system's personnel could offer their own opinions.''} (NP5) 
& - 
& - 
& \hlcell[50]{np}{7/16}  \\
\hline
\end{tabular}
\end{table*}

\subsection{Perceived Workload (RQ1)}
\label{sec: NASA-TLX}

Participants’ perceived workload was assessed using the NASA-TLX.
Table~\ref{tab:mental_demand} presents the mean and standard deviation for each workload metric across the three conditions (\condN, \condS, \condM), along with the results of the Kruskal–Wallis tests.
We did not find any significant difference across the interface conditions for any of the NASA-TLX dimensions.

\subsection{Interaction Behavior Analysis (RQ2)}
\label{sec: interaction-behavior}

We further analyzed how frequently the participants performed \textit{Annotations} (e.g., highlighting and commenting) and \textit{Reactions} (e.g., pinning agents, checking references) under the \conditionS and the \conditionM.
Figure \ref{fig:interaction_log1} shows the scatter plot of the average counts of \textit{Annotations} and \textit{Reactions} per participant.
We applied Silhouette analysis to the scatter plots of \conditionS and \condM 
after excluding data points where both \textit{Annotations} and \textit{Reactions} were zero.
The resulting Silhouette score was 0.38, indicating a moderate separation between the conditions.
These results suggest that, although there is some overlap, participants in \condM and \conditionS tended to exhibit distinguishable interaction patterns.
Our data showed that 12 out of 15 participants in the \conditionM never used \textit{Annotations}, which is 3 times higher than in \conditionS (4/15).
In contrast, participants in the \conditionM used \textit{Reactions} more frequently than those in the single condition.
In \conditionS, participants primarily performed \textit{Annotation} interactions while \textit{Reaction} interactions were dominant in \conditionM.

\subsection{Open-Ended Question Analysis (RQ2)}
\label{sec:qualitative}

We further examined the open-ended responses to understand the user experience of the \conditionS and \conditionM.
Table \ref{tab:themes} summarizes the derived themes from our analysis.

\subsubsection{Perceived Skill Change when Receiving Different Types of Feedback}
We found that participants in the \conditionN more frequently mentioned improvements in lower-level thinking skills (\textit{Knowledge}, \textit{Comprehension}, and \textit{Application})
as well as enabling skills (reading speed, and concentration),
as one participant from \conditionN mentioned \textit{``reading speed has improved''} and \textit{``read more attentively''} (NP2).

In contrast, participants in the \condM and \conditionS more often reported improvements in both higher-level thinking skills (\textit{Analysis}, \textit{Synthesis}, and \textit{Evaluation}) and lower-level thinking skills,
as one participant from \conditionM mentioned \textit{``I read through the suggestions and comments from AIs designed as specialists, which evaluate information relevant to my answers and could be incorporated to improve their quality or clarity''} (MP8).
This difference may help explain why the \condM and \conditionS outperformed the \conditionN in terms of critical thinking ability improvements. 

More participants in the \conditionM mentioned improvements in their analysis skills (8/15) compared to the \conditionS (2/15): \textit{``My reading skills, particularly in terms of analyzing research papers, has improved considerably''} (MP2). 

\subsubsection{Thought Exchange with AI Agents}
A majority of participants highlighted the value of exchanging thoughts.
Specifically, over two-thirds of participants in both the \condS and \conditionM (12/15 and 11/15, respectively)
reported that thought exchange with AI agents was the most useful feature of the system.
Furthermore, nine participants in \conditionM emphasized that the feature of multiple agents helped them understand the paper from different angles.
For example, one participant mentioned, \textit{``I think the system’s AI feature that analyzes my answers is the most helpful... the three AI assistants can identify my limitations from different angles based on my responses. This allows me to gain new insights from each practice''} (MP10).
This suggests that exposing learners to different disciplinary viewpoints can enhance their engagement and reflection during critical reading tasks. 

In contrast, responses from the \conditionN were more dispersed. 
Seven participants from \conditionN complained that the fixed feedback is the least helpful or satisfying feature.
But four participants rather expressed positive opinions on the fixed feedback, mentioning that it helped them reflect on their answers. 
Since self-reflection is known to facilitate self-regulated learning~\cite{Zimmerman2002Becoming},
this might explain why participants in the \conditionN were able to approach the critical thinking questions from more viewpoints after receiving the fixed feedback (see Section \ref{sec: viewpoints}).

\subsubsection{Attitudes toward Feedback Given By AI Agents}
While thought exchange with multiple AI agents increased the amount of information available to participants, it also tended to be formulaic,
which sometimes distracted attention and undermined confidence.
This effect was specific to the multi-agent condition because the agents sometimes produced similarly structured responses, which made the interaction appear repetitive.
A few participants reported issues with the AI agent’s feedback.
In the \conditionM, four participants described the feedback as nitpicking and formulaic, noting that all feedback was given in a \textit{``Yes, but…''} (MP3) format, which sometimes undermined their confidence in answering the questions. 
Three participants in the \conditionM reported experiencing contextual drift, stating that the feedback included information outside the current text in the paper (MP13) or their focus (MP1).
Only one participants in the \conditionS mentioned the contextual drift, and no participants reported the feedback as nitpicking even though we used the same LLM prompt.


\section{Discussion}

\subsection{Reflections on the Research Questions}
\subsubsection{Incorporating Agent(s) Feedback Promotes Critical Thinking (RQ1)}
The user study results suggest that involving thought exchanges with AI agents in paper reading (\condM and \conditionS) helped our participants better practice their critical thinking skills (Section \ref{sec: external review}), improved their self-perceived critical thinking abilities (Section \ref{sec: self critical thinking}), and assisted them in generating more distinct viewpoints (Section \ref{sec: viewpoints}).
Importantly, the results highlight a benefit of in-situ thought exchange, which offers substantial support specifically for junior researchers, aligning with one of our central motivations.

These findings provide an example of how LLMs can be critically utilized in learning tasks, especially in the context where many researchers are concerned that over-reliance on AI may have negative side effects on people’s critical thinking skills~\cite{gerlich2025ai}. 
Our study demonstrates that integrating LLMs into the academic paper reading process by facilitating thought exchange can enhance junior researchers’ critical thinking abilities. 
Most existing LLM-assisted tools on academic practice mainly focus on improving ``efficiency'' for researchers to quickly retrieve information or synthesize the literature~\cite{Chi2005ScentHighlights, Fok2023Scim, Fok2024Accelerating, Huth2024Eye, Gu2024An}.
However, this kind of cognitive offloading approach may undermine people's higher-level thinking skills~\cite{carr2020shallows, sparrow2011google}.
Prior work on bot-facilitated online deliberation and discussion systems suggests that conversational exchanges with agents can scaffold perspective-taking~\cite{SeeWidely2024}, and argument evaluation~\cite{Wambsganss2021ArgueTutor}. 
Such dialog engagement is distinct from passive consumption of AI-generated output and aligns more closely with the processes essential for critical thinking.
Building on this insight, our design shifts the role of AI from cognition delegation to reflective interlocutor.
When an agent challenges the users' response, such as suggesting the answer is not sufficiently broad or lacks depth, it can elicit one of two productive outcomes: the user accepts the feedback and revises their thinking, or they question the agent's claim, which constitutes a form of critical reflection.

These findings also align with previous research on group reading, which has shown that pre-reading a text and organizing group reading sessions to exchange diverse perspectives can enhance readers’ comprehension and critical thinking skills~\cite{kamin2002does,perry1999forms}.
Our results suggest that LLM-enabled agents may effectively serve a similar role, simulating the presence of peers and prompting users to reflect on different perspectives during reading.
In future work, the same approach could be applied to situations where collaboration is needed but opportunities for interaction are limited, such as remote asynchronous learning environments, cross-cultural and cross-disciplinary collaborations.

Our results did not reveal significant differences in subjective workload across the three conditions (Section \ref{sec: NASA-TLX}).
We did not find evidence suggesting that interacting with agents in a paper-reading task significantly added extra workload to participants. One possible explanation is that paper-reading was already a cognitively demanding task. Having additional interaction with AI agents in the ongoing task may not have introduced perceivable differences in workload.
Future studies should further investigate how in-situ interface designs can minimize cognitive overheads for thought exchanges in paper reading.

\subsubsection{Effects of Single vs. Multiple AI Agents on Interaction Strategies (RQ2)}

The difference in interaction behaviors (Section \ref{sec: interaction-behavior}) uncovered that the thought exchange setups affected participants' reading behavior.
In our study, participants in the \conditionM primarily focused on analyzing answers provided by multiple AI agents.
In contrast, participants in the \conditionS more frequently integrated comments by AI agents with the content in a given paper.
This suggests that future interfaces may adopt various setups to engage users in different types of cognitive activities during reading.
For example, 
A setup with multiple AI agents may better help users identify relevant information beyond a given paper, whereas a one-on-one setup is more suitable for users to focus on and consolidate their understanding of the paper's content.

Although we did not observe clear quantitative differences in effectiveness between \condS and \conditionM, our open-ended question analysis results uncovered participants' perception of learning analysis skills in \conditionM. 
As shown in Table \ref{tab:themes}, 8 out of 15 participants in \conditionM reported improvements in their \textit{Analysis} abilities, four times more than those in \conditionS.
\textit{Analysis} is a core higher-order cognitive process in Bloom’s Taxonomy~\cite{bloom1956taxonomy}, involving the ability to break down information and compare viewpoints, which is essential for critical thinking.
A possible explanation is that engaging with multiple agents in parallel and using our proposed interface to add reactions such as \textit{pin}, \textit{thumb up}, or \textit{thumb down} (\autoref{fig:system_intro}-C2, C3) required participants to compare and analyze different comments.
This feature illustrates a unique advantage of multi-agent thought exchange interfaces, which can prompt users to engage in comparative and analytical reasoning.
From the theoretical perspective, this behavior can be interpreted as a form of metacognition scaffolding~\cite{flavell1979metacognition}, in which external prompts encourage readers to monitor and compare their own reasoning  during reading process. 

Our analysis of the responses to the post-experimental open-ended questions (Section \ref{sec:qualitative}) revealed shortcomings and current implementation limitations in \conditionM.
Participants mentioned they considered some comments from AI agents in \conditionM to be rather nitpicking or deviating from their current focus. 

One possible reason might be that the prompts (Appendix \ref{Appendix: Prompts}) for all AI agents in the \textit{Section Pane} (Section \ref{sec: Section Pane}) followed the same format:
they first evaluated the reader's answer and then provided perspectives from different academic disciplines. 
To enhance the diversity of feedback, we designed the agents to emphasize knowledge from distinct domains, which may sometimes fall outside the users' interests.
This issue may lead to a negative user experience; however, our design still effectively exposed users to diverse perspectives. 
Future work should allow users to select their preferred persona, such as encouragement-focused agents without criticism or agents providing more direct feedback.
Additionally, further exploration is encouraged to provide users with options to choose which academic disciplines the agents represent, allowing them to interact with agents most relevant to their interests.

On the contrary, participants in \conditionS reported fewer negative expressions to the AI agent, despite the feedback formulation and prompts being identical, with the only difference being the number of agents.
This suggests that the number of agents engaged in the thought exchange can influence user experience.
Under cognitively demanding tasks, such as critical paper reading, involving multiple agents may elicit more negative feelings.
In contrast, comments from a single agent are potentially easier to process, as users do not need to continually adapt to different domains, styles, or reasoning approaches.
This observation is consistent with prior work showing that multi-agent systems do not necessarily outperform single-agent systems~\cite{pan2025why}.
Future work should explore strategies to balance the benefits of multiple agents with the need to avoid overwhelming users. 
Approaches could include gradually introducing multiple perspectives based on task difficulty. 
Such designs may help preserve the diversity of insights while keeping cognitive load manageable.

\subsection{Limitations and perspective for future work}
While our studies provide valuable insights into how interacting with a single and multi-agent interface benefits critical reading, there are also limitations. 

First, we only simulated agents representing different research domains with LLM to generate a diverse set of thoughts.
Simulated AI agents are not necessarily fully equivalent to human experts in terms of knowledge and thinking capacity.
In addition, the responses generated by these AI agents may overlap because the controls by LLM prompts are not strict.
Relatedly, our AI agent design remains limited in that agents operate largely in parallel and independently, rather than engaging in an interactive scholarly dialogue. 
Future work should investigate more algorithmic and interaction-driven designs to generate diverse thoughts and allow agents to respond to and build on one another.
Moreover, the agent mediated conditions differed not only in agent count but also in domain assignment, the \textsc{multi condition} used dynamically selected disciplines, whereas the \textsc{single condition} relied on a fixed domain outside participants’ expertise, which introduces a potential confound.
Future work could address this by ensuring that participants across different conditions encounter domains matched in familiarity, difficulty, and number, in order to better understand how variations in agent expertise contribute to participants’ critical thinking outcomes.

Second, the present user study
focused on a relatively small and specific sample of participants. 
Future studies could include a larger and more diverse participant pool to examine whether the patterns we observed hold across different levels of expertise or reading strategies.
It is also worth exploring the long-term development of critical thinking skills following the study, without interface support, to better capture reading and critique behaviors without explicit scaffolding.

Third, LLM-based agents may generate hallucinated statements or adopt incorrect disciplinary framings. To mitigate this risk, we designed a reference feature for each agent-generated statement to encourage readers to verify claims against the source paper.
However, it is inevitable that hallucinations cannot be fully prevented in LLMs.
Rather than trying to eliminate hallucinations entirely, future work could explore interaction designs that help users rely on AI outputs more appropriately and think more critically about agent responses.

Moreover, it would be valuable and interesting for future work to investigate the reasons behind participants’ interaction strategies through interviews. For example, such studies could explore why participants in \conditionM rarely utilized annotations, which may reveal interesting insights about how multi-agent feedback is processed and integrated.


\section{Conclusion}

In this work, we introduced an LLM-based in-situ thought exchange interface that integrates simulated peer feedback to augment critical reading.
Our mixed-methods user study demonstrated that (1) incorporating agent-mediated thought exchanges during academic paper reading fostered junior researchers’ critical thinking skills; and (2) the number of agents involved influenced readers’ critical reading practices in distinct ways. 
Specifically, 
participants interacting with a single agent tended to annotate directly in the text after viewing the AI agent’s response as a way to reflect on their thoughts; 
in contrast, those engaging with multiple agents primarily focused on comparing and analyzing the different perspectives expressed by the AI agents, while paying less attention to annotating directly on the text.
Our contribution lies in designing LLMs that encourage users to invest more cognitive effort into reading instead of offloading it.

\section{GenAI Usage Disclosure}
In this work, we utilized ChatGPT (GPT-4o) and Writefull to assist with editing and refining portions of the manuscript text for clarity and language fluency.
Additionally, AI tools were used to assist in code formatting and minor scripting tasks, 
but the authors performed all data analysis and code development.
All scientific ideas, analyses, experimental designs, and interpretations of results were solely developed by the authors.

\begin{acks}
We thank Zefan Sramek, Shitao Fang, Jo Takezawa, and the other members of IIS Lab for their support during this work. This work was partly supported by JST ASPIRE for Top Scientists (Grant Number JPMJAP2405), JST PRESTO (Grant Number JPMJPR23IB), JST SPRING (Grant Number JPMJSP2108) and the JSPS Invitational Fellowship for Research in Japan program (Long-term, Grant Number L24509).
\end{acks}


\bibliographystyle{ACM-Reference-Format}
\bibliography{9-Reference}

\appendix
\onecolumn
\section{Formative Study Participants}
\label{appendix: formative study}
\begin{table}[htp]
\centering
\caption{Participant Information}
\renewcommand{\arraystretch}{1.2}
\begin{tabular}{c|c|c|c|c|c}
\hline
\textbf{ID} & \textbf{Age} & \textbf{Gender} & \textbf{Years of Experience} & \textbf{Current Role} & \textbf{Major} \\
\hline
\hline
P1  & 27 & M & 5  & PhD Student & Chemical System Engineering         \\
P2  & 28 & M & 6  & PhD Student     & Computer Vision                \\
P3  & 23 & M & 2.5  & Master Student     & HCI     \\
P4  & 25 & M & 4  & PhD Student     & Computer Vision      \\
P5  & 24 & M & 2  & Master Student     & HCI            \\
P6  & 28 & M & 6  & PhD Student     & Economics               \\
P7  & 26 & M & 5  & PhD Student     & HCI \\
P8  & 23 & F & 1  & Master Student  & Media               \\
\hline
\end{tabular}

\label{tab:participant_info}
\end{table}
\section{Critical Thinking Self-Assessment Questionnaire}
\label{ref:appendixA}
Please indicate your level of agreement with the following statement regarding the system. (7-point Likert scale: 1 = strongly disagree, 7 = strongly agree; Note that all italicized texts do not appear in the questionnaire.)
\begin{enumerate}
    \item \textit{Interpretation}
    \begin{enumerate}
        \item The system can help me restate another perspective's statements to clarify the meaning.
        \item The system helps me to seek clarification of the meanings of another perspective’s opinion or points of view.
    \end{enumerate}

    \item \textit{Analysis}
    \begin{enumerate}
        \item With the system's help, I examine the interrelationships among concepts or opinions posed.
        \item The system helps me to figure out if the author’s arguments include both for and against the claim.
    \end{enumerate}

    \item \textit{Evaluation}
    \begin{enumerate}
        \item I assess the contextual relevance of opinions or claims posed by the system.
        \item The system helps me to search for additional information that might support or weaken an argument.
    \end{enumerate}

    \item \textit{Inference}
    \begin{enumerate}
        \item The system helps me seek useful information to refute an argument when supported by unclear reasons.
        \item The system helps me systematically analyse the problem using multiple sources of information to draw inferences.
    \end{enumerate}

    \item \textit{Explanation}
    \begin{enumerate}
        \item I can explain a key concept to clarify my thinking with the system's help.
        \item I present more evidence or counter evidence for another perspective's points of view with the system's help.
    \end{enumerate}

    \item \textit{Self-regulation}
    \begin{enumerate}
        \item The system helps me revise and rethink strategies to improve my thinking.
        \item The system helps me reflect on my thinking to improve the quality of my judgment.
    \end{enumerate}
\end{enumerate}

\section{Late-breaking Work Paper Used in the User Study}
\label{ref:appendix_LBW}
Note that Paper 1 and Paper 8 were swapped for half of the participants to counterbalance the sequence order.

\begin{table*}[htp]
\centering
\renewcommand{\arraystretch}{1.3} 
\setlength{\tabcolsep}{12pt}      
\rowcolors{2}{gray!15}{white}     

\begin{tabular}{>{\centering\arraybackslash}m{2.5cm} >{\centering\arraybackslash}m{8cm}}
\toprule
\textbf{Paper ID} & \textbf{Reference} \\
\midrule
Paper 1 & Mhasakar et al. \cite{Mhasakar2025I} \\
Paper 2 & Kruk et al. \cite{Kruk2025BanglAssist} \\
Paper 3 & Shukla et al. \cite{Shukla2025De-skilling} \\
Paper 4 & Liu et al. \cite{Liu2025Can} \\
Paper 5 & Cifliku et al. \cite{Cifliku2025This} \\
Paper 6 & Lee et al. \cite{Lee2025MAP} \\
Paper 7 & Huang et al. \cite{Huang2025Vipera} \\
Paper 8 & Gadhvi et al. \cite{Gadhvi2025AdaptAI} \\
\bottomrule
\end{tabular}
\caption{Paper ID and Reference.}
\label{tab:chi_papers}
\end{table*}

\section{Prompts}
\label{Appendix: Prompts}
Note that \condN, \condS, and \conditionM share the same prompt for critical thinking questions generation.
In \conditionS, the prompt to generate single-disciplinary feedback differs slightly from that in \conditionM: it provides feedback from one disciplinary perspective that is different from the participant’s major, and without \{pinned agent\} input.
The text inside \{\} represents the participants' input.

\subsection{Prompt for Critical Thinking Questions Generation}
\lstset{
  backgroundcolor=\color{gray!20}, 
  basicstyle=\ttfamily\footnotesize, 
  breaklines=true,                 
  frame=single,                    
  framerule=0pt,                   
  xleftmargin=5pt, xrightmargin=5pt 
}
\begin{lstlisting}
### **Input**  
Section: {text}  
Category: {category}

### **Output**  
- Question  

### **Requirement**
1. Read the given section content carefully.
2. Identify the main topic and the core ideas.
3. Generate **only one** critical thinking question that is directly relevant to the section and aligned with the assigned category.
4. Ensure that the question follows the structure and tone of the example questions provided.
5. Return **only the question** with no extra formatting or explanation.
6. Ensure the question is:
    - Clear and easy to understand for junior researchers, using simple and precise language.
    - Logically structured and specific, avoiding vague or overly abstract expressions.
    - Thought-provoking and rigorous, encouraging the reader to reflect critically.

### **Examples**  
**Input**  
Section: {text}  
Category: {category}  

**Output**  
{few_shot_questions}
\end{lstlisting}








\subsection{Prompt for Multi-Disciplinary Agent Feedback}
\begin{lstlisting}
The instructor asks a critical thinking question '{question}', and the reader's answer is '{answer}'. 
Judge the reader's answer first, then provide an improved response from **experts in three distinct and relevant academic disciplines**. 
First generate answers from '{concatenated_disciplines}'s perspectives. 
Each answer should emphasise how it **aligns with or challenges current trends or theories** in that discipline. 
Ensure that each discipline's answer introduces fundamentally different insights rather than paraphrasing the same idea in different terminology.
Each answer should:  
- **Begin with an implicit evaluation** of the reader's answer before transitioning into the improved response.  
- **Highlight alignment with or challenges to current trends or theories** in that discipline.  
- **Ensure each discipline's answer introduces fundamentally different insights**, avoiding rephrasing the same idea.  
- **Be concise (less than 100 words per answer).**  

In the output, replace any instance of 'The reader's answer' with 'Your answer'.
Provide the output as **raw JSON text**, without any surrounding Markdown formatting like triple backticks.
Do not split or abbreviate any disciplines. Each discipline in '{concatenated_disciplines}' should be treated as a distinct entity. 
If fewer than three disciplines are provided, choose an additional relevant discipline to complete the set of three.
    {
      [Perspective Name]: [Judgement of answer][Improved answer],
      [Perspective Name]: [Judgement of answer][Improved answer],
      [Perspective Name]: [Judgement of answer][Improved answer]
    }
\end{lstlisting}

\subsection{Prompt for Reinterpreting Highlights via Multi-Disciplinary Agents}
\begin{lstlisting}
The reader highlights '{quote}' in the paper and marks it as '{label}'. 
Reinterpret '{label}' of reader's highlight from ** 3 different academic disciplines's perspective**. 
First generate reinterpretation from '{concatenated_disciplines}'s perspectives. 
Each reinterpretation should emphasise how it **aligns with or challenges current trends or theories** in that discipline. 
Ensure that each discipline's reinterpretation introduces fundamentally different insights rather than paraphrasing the same idea in different terminology.
Each reinterpretation should:  
- **Highlight alignment with or challenges to current trends or theories** in that discipline.  
- **Ensure each discipline's reinterpretation introduces fundamentally different insights**, avoiding rephrasing the same idea.  
- **Be concise (less than 100 words per reinterpretation).**  

Provide the output as **raw JSON text**, without any surrounding Markdown formatting like triple backticks.
Do not split or abbreviate any disciplines. Each discipline in '{concatenated_disciplines}' should be treated as a distinct entity. 
If fewer than three disciplines are provided, choose an additional relevant discipline to complete the set of three.
    {
        [Perspective Name]: [Reinterpretation],
        [Perspective Name]: [Reinterpretation],
        [Perspective Name]: [Reinterpretation]
    }    
\end{lstlisting}

\subsection{Prompt for Answering Comments via Multi-Disciplinary Agents}
\begin{lstlisting}
The reader highlights '{quote}' in the paper and adds comment: '{comment}'. 
If the reader asks a question, answer the question based on highlights from ** 3 most relevant different academic disciplines's perspective**. 
If the reader leaves a statement, discuss around reader's highlight and comment from ** 3 most relevant different academic disciplines's perspective**. 
First generate a response from '{pinned_agent}'s perspectives. 
Each response should emphasize how it **aligns with or challenges current trends or theories** in that discipline. 
Ensure that each discipline's response introduces fundamentally different insights rather than paraphrasing the same idea in different terminology.
Each response should:  
- **Highlight alignment with or challenges to current trends or theories** in that discipline.  
- **Ensure each discipline's response introduces fundamentally different insights**, avoiding rephrasing the same idea.  
- **Be concise (less than 100 words per response).**  

Provide the output as **raw JSON text**, without any surrounding Markdown formatting like triple backticks.
Do not split or abbreviate any disciplines. Each discipline in '{pinned_agent}' should be treated as a distinct entity. 
If fewer than three disciplines are provided, choose an additional relevant discipline to complete the set of three.
   {
      [Perspective Name]: [Response],
      [Perspective Name]: [Response],
      [Perspective Name]: [Response]
    }
\end{lstlisting}

\subsection{Prompt for Revised Answer Feedback}
\begin{lstlisting}
User's initial answer is '{initialAnswer}', the revised answer is '{revisedAnswer}'. 
The feedback for the initial answer is '{agreeText}'. 
Based on this feedback, evaluate how well the revised answer addresses the issues or suggestions mentioned. 
Provide a concise and constructive feedback on the revised answer in 100 words.
\end{lstlisting}


\section{Static Feedback in \conditionN}
\label{appendix: static feedback}
\begin{itemize}
    \item Did your answer accurately reflect what is stated in the paper?
    \item Did you miss any important information, details, or alternative perspectives?
    \item Really stuck? You can re-read the section and try again!
    \item Take a moment to reflect on your answer—what could you improve or clarify? Try to answer it again.
\end{itemize}
\section{Details of User Study Result}
\label{Appendix: User Study Result}
\begin{table*}[htbp]
\centering
\scriptsize
\caption{Participants’ response quality scores and  statistical effects in pre- and post-training test}
\begin{tabular}{c | c c c | l| l| l}
\hline
Measure & Condtion & Post (Mean $\pm$ SD) & Pre (Mean $\pm$ SD) & Condition Effect & Stage Effect & Condition $\times$ Stage \\
\hline
\multirow{3}{*}{Total} & \hlcell[100]{np}{\textsc{\condN}}   & 2.72 $\pm$ 0.58 & 2.56 $\pm$ 0.64 & $F(2,43)=1.34$ & $F(1,43)=43.32$ & $F(2,43)=4.94$ \\
& \hlcell[100]{sp}{\textsc{\condS}}  & 3.15 $\pm$ 0.35 & 2.48 $\pm$ 0.64 & $p=.27$ & $p<.001^{***}$ & $p<.05^{*}$ \\
& \hlcell[100]{mp}{\textsc{\condM}}  & 3.17 $\pm$ 0.42 & 2.74 $\pm$ 0.55 & $ \eta_p^2=.06$ & $ \eta_p^2=.50$ & $\eta_p^2 = .19$ \\
\hline

\multirow{3}{*}{\makecell{Explanation\\ of issues}}
 & \hlcell[100]{np}{\textsc{\condN}}   & 3.05 $\pm$ 0.58 & 2.69 $\pm$ 0.69 & $F(2,43)=1.77$ & $F(1,43)=49.18$ & $F(2,43)=2.29$ \\
& \hlcell[100]{sp}{\textsc{\condS}}  & 3.45 $\pm$ 0.37 & 2.75 $\pm$ 0.65 & $p=.18$ & $p<.001^{***}$ & $p=.11$ \\
& \hlcell[100]{mp}{\textsc{\condM}}  & 3.45 $\pm$ 0.37 & 3 $\pm$ 0.52 & $ \eta_p^2=.08$ & $ \eta_p^2=.53$ & $\eta_p^2 = .10$ \\
\hline

\multirow{3}{*}{Evidence}
 & \hlcell[100]{np}{\textsc{\condN}}   & 2.73 $\pm$ 0.69 & 2.56 $\pm$ 0.74 & $F(2,43)=1.61$ & $F(1,43)=31.32$ & $F(2,43)=4.58$ \\
& \hlcell[100]{sp}{\textsc{\condS}}  & 3.28 $\pm$ 0.35 & 2.5 $\pm$ 0.71 & $p=.21$ & $p<.001^{***}$ & $p<.05^{*}$ \\
& \hlcell[100]{mp}{\textsc{\condM}}  & 3.28 $\pm$ 0.47 & 2.78 $\pm$ 0.65 & $ \eta_p^2=.07$ & $ \eta_p^2=.42$ & $\eta_p^2 = .18$ \\
\hline

\multirow{3}{*}{\makecell{Influence of \\ context and \\ assumptions}}
 & \hlcell[100]{np}{\textsc{\condN}}   & 2.61 $\pm$ 0.58 & 2.47 $\pm$ 0.61 & $F(2,43)=1.53$ & $F(1,43)=31.25$ & $F(2,43)=2.82$ \\
& \hlcell[100]{sp}{\textsc{\condS}}  & 2.97 $\pm$ 0.49 & 2.4 $\pm$ 0.63 & $p=.23$ & $p<.001^{***}$ & $p=.07$ \\
& \hlcell[100]{mp}{\textsc{\condM}}  & 3.08 $\pm$ 0.58 & 2.63 $\pm$ 0.52 & $ \eta_p^2=.07$ & $ \eta_p^2=.42$ & $\eta_p^2 = .12$ \\
\hline

\multirow{3}{*}{\makecell{Student's position}}
 & \hlcell[100]{np}{\textsc{\condN}}   & 2.66 $\pm$ 0.58 & 2.48 $\pm$ 0.64 & $F(2,43)=1.03$ & $F(1,43)=39.39$ & $F(2,43)=3.78$ \\
& \hlcell[100]{sp}{\textsc{\condS}}  & 3.02 $\pm$ 0.42 & 2.28 $\pm$ 0.67 & $p=.37$ & $p<.001^{***}$ & $p<.05^{*}$ \\
& \hlcell[100]{mp}{\textsc{\condM}}  & 3.07 $\pm$ 0.48 & 2.6 $\pm$ 0.55 & $ \eta_p^2=.05$ & $ \eta_p^2=.48$ & $\eta_p^2 = .15$ \\
\hline

\multirow{3}{*}{\makecell{Conclusion and \\ related outcomes}}
 & \hlcell[100]{np}{\textsc{\condN}}   & 2.55 $\pm$ 0.63 & 2.59 $\pm$ 0.64 & $F(2,43)=0.88$ & $F(1,43)=11.04$ & $F(2,43)=6.95$ \\
& \hlcell[100]{sp}{\textsc{\condS}}  & 3.03 $\pm$ 0.36 & 2.47 $\pm$ 0.7 & $p=.42$ & $p<.01^{**}$ & $p<.01^{**}$ \\
& \hlcell[100]{mp}{\textsc{\condM}}  & 2.98 $\pm$ 0.42 & 2.69 $\pm$ 0.67 & $ \eta_p^2=.04$ & $ \eta_p^2=.20$ & $\eta_p^2 = .24$ \\
\hline

\end{tabular}
\end{table*}


\end{document}